\documentclass[fleqn,usenatbib]{mnras}

\usepackage{newtxtext,newtxmath}

\usepackage[T1]{fontenc}

\DeclareRobustCommand{\VAN}[3]{#2}
\let\VANthebibliography\thebibliography
\def\thebibliography{\DeclareRobustCommand{\VAN}[3]{##3}\VANthebibliography}

\usepackage{graphicx}	
\usepackage{amsmath}	

\newcommand{\mps}{\; {\rm m\;s^{-1}}}
\newcommand{\mpss}{\; {\rm m^2\;s^{-2}}}

\newcommand{\btt}{}

\def \tdrag {\tau_\mathrm{drag}}

\title[Atmospheric circulation of brown dwarfs and directly imaged exoplanets]{Atmospheric circulation of brown dwarfs and directly imaged exoplanets driven by cloud radiative feedback: effects of rotation}

\author[Tan \& Showman]{
Xianyu Tan$^{1}$\thanks{E-mail: xianyu.tan@physics.ox.ac.uk} and
Adam P. Showman$^{2,3,\dagger}$
\\
$^{1}$Atmospheric, Oceanic  and Planetary Physics, Department of Physics, University of Oxford, OX1 3PU, UK\\
$^{2}$Lunar and Planetary Laboratory, University of Arizona, 1629 University Boulevard, Tucson, AZ 85721, USA \\
$^{3}$Department of Atmospheric and Oceanic Sciences, Peking University, Beijing, People’s Republic of China\\
$^{\dagger}{\rm Deceased.}$
}

\date{Accepted XXX. Received YYY; in original form ZZZ}

\pubyear{2021}

\begin{document}
\label{firstpage}
\pagerange{\pageref{firstpage}--\pageref{lastpage}}
\maketitle

\begin{abstract}
 Observations of brown dwarfs (BDs), free-floating planetary-mass objects and directly imaged extrasolar giant planets (EGPs) exhibit rich evidence of large-scale weather. Cloud radiative feedback has been proposed as a potential mechanism driving the vigorous atmospheric circulation on BDs and directly imaged EGPs, and  yet it has not been demonstrated in  three-dimensional (3D) dynamical models at relevant conditions. Here we present a series of atmospheric circulation models that self-consistently couple dynamics with idealized cloud formation and its radiative effects. We demonstrate  that vigorous atmospheric circulation can be triggered and self-maintained by  cloud radiative feedback. Typical isobaric temperature variation could reach over 100 K and horizontally averaged wind speed could be several hundred $\mps$.  The circulation is dominated by cloud-forming and clear-sky vortices that evolve over timescales from several to tens of hours. The typical horizontal lengthscale of dominant vortices is closed to the Rossby deformation radius, showing a linear dependence on the inverse of rotation rate. Stronger rotation tends to weaken vertical transport of vapor and clouds, leading to overall thinner clouds. Domain-mean outgoing radiative flux exhibits variability over timescales of tens of hours due to the statistical evolution of storms. Different bottom boundary conditions in the models could lead to qualitatively different circulation near the observable layer.   The circulation driven by cloud radiative feedback represents a robust mechanism generating significant surface inhomogeneity as well as irregular flux time variability. Our results  have important implications for near-IR colors of dusty BDs and EGPs, including the scatter in the near-IR color-magnitude diagram and the viewing-geometry dependent  near-IR colors.
\end{abstract}




\section{Introduction}
\label{ch.intro}

Active weather is likely common among brown dwarfs (BDs) as has been indicated by several lines of evidence. First, atmospheric circulation drives  temperature anomalies and inhomogeneous cloud coverage on a global scale, which is likely responsible for the observed lightcurve variability of many L and T dwarfs (e.g., \citealp{artigau2009, radigan2012,buenzli2012, apai2013,buenzli2014,wilson2014, metchev2015,  yang2016,leggett2016,MilesPaez2017, apai2017,manjavacas2017,zhou2018,vos2019,eriksson2019,lew2020,zhou2020,bowler2020,Hitchcock2020,vos2020}, see also recent reviews by \citealp{biller2017} and \citealp{artigau2018}). Second, spectra and near-IR colors of many L dwarfs suggest the presence of thick clouds in their photospheres (e.g., \citealp{chabrier2000, allard2001,tsuji2002, burrows2006, saumon2008, charnay2018}). The existence of thick clouds indicates the presence of atmospheric dynamics against gravitational settling. Third,  the abrupt transition from L to T dwarfs is likely caused by either a sudden change of cloud patchiness or thickness  \citep{ackerman2001,burgasser2002, knapp2004,marley2010}, for which atmospheric dynamics likely plays a role. Fourth, large-scale circulation provides a source of vertical mixing that helps to explain the inferred chemical disequilibrium in a wide  range of  BDs (e.g., \citealp{saumon2006,stephens2009,leggett2016b,leggett2019, miles2020}), especially in stratified atmospheres where convection does not play a direct role in mixing. Other techniques detecting or constraining the presence of global circulation of BDs include Doppler imaging \citep{crossfield2014}, simultaneous tracking of near-IR and radio variability \citep{allers2020} and precise near-IR polarization measurements \citep{millar2020}. 

Extrasolar giant planets (EGPs)  detected by the direct imaging technique so far are mostly young, hot and relatively distant from their host stars, and therefore they receive negligible stellar bolometric flux compared to their internal heat flux. Their atmospheric structure and dynamics are likely determined  mostly by internal luminosity, and thus they fall into the same category {\btt as} field BDs in terms of atmospheric characteristics. Similar to BDs, the near-IR spectrum and color of most directly imaged EGPs suggest the presence of thick clouds and possibly significant chemical disequilibrium in their photospheres (e.g., \citealp{currie2011,  barman2011a,barman2011b, marley2012,   oppenheimer2013, ingraham2014,rajan2015, moses2016,chauvin2017, stolker2020}). Although current instruments place a high threshold  for the detection of lightcurve variability of directly imaged EGPs orbiting bright stars\footnote{\btt Detecting variability of directly imaged exoplanets requires precisely removing contamination from the point spread function of the bright hot star. The current state-of-the-art ground based extreme adaptive optics observations can reach $\sim$10\% precision in high-cadence time series \citep{apai2016}. Typical amplitudes of rotational modulations are much smaller than this level \citep{artigau2018}. Additional limitations may arise if the viewing angle of the exoplanets is close to polar view (and therefore the rotational modulation is minimized), or due to  stellar variability.}, their  planetary-mass, free-floating   counterparts commonly exhibit lightcurve variability \citep{biller2015,zhou2016,vos2018,biller2018,schneider2018,manjavacas2019,miles2019}.


Understanding the atmospheric circulation of BDs and directly imaged EGPs is  a  pressing need given the large body of observational constraints. Unlike planets in close-to-modest distances from their host stars whose atmospheric circulation is primarily driven by the  horizontal differential stellar irradiation, atmospheres of relatively isolated BDs and directly imaged EGPs lack  horizontal differential heating from external sources, and their atmospheric circulation is driven almost solely by internal heat flux. So far, two major categories of sources have been proposed to drive the global circulation in the stratified, observable atmospheric layers. 

The first is a mechanically forced scenario: convection interacts with the overlying stratified atmospheres and generates a wealth of waves and turbulence. These atmospheric eddies  interact with the large-scale flows, driving  large-scale circulation. There have been a few studies in this direction.  Local hydrodynamic simulations by \cite{freytag2010}  show that gravity waves generated by interactions between the convective interior and the stratified layer can cause mixing and lead to small-scale cloud patchiness.    \cite{showman&kaspi2013} presented global convection models and analytically estimated typical wind speed and horizontal temperature differences driven by the absorption and breaking of atmospheric waves in the stably stratified atmosphere.  By injecting random  forcing to a shallow-water system,  \cite{zhang&showman2014} showed that weak radiative dissipation and strong forcing  favor large-scale zonal jets for BDs, whereas strong dissipation and weak forcing favor transient eddies and quasi-isotropic turbulence. Using a general circulation model coupled with parameterized thermal perturbations resulting from interactions between convective interior and the stratified atmosphere,  \cite{showman2019} showed that under conditions of relatively strong forcing and weak damping, robust zonal jets and the associated meridional circulation and temperature structure are common outcomes of the dynamics. They also demonstrated that  long-term (multi months to years) quasi-periodic oscillations on the equatorial zonal jets, similar to the Quasi-biennial oscillation observed in Earth's stratosphere, can be driven by the thermal perturbations.

The second scenario is a thermally-driven mechanism linked to cloud radiative feedback \citep{tan2019}. Clouds are critical in shaping the thermal structure, near-IR color and spectral properties of substellar atmospheres via large opacity loading (see recent reviews by \citealp{helling2014}, \citealp{marley2015} and \citealp{helling2019}).  Cloud radiative feedback is similarly important in driving a vigorous global circulation and atmospheric variability.  Imagine an atmosphere consisted of patchy clouds. In the cloudy region, less thermal radiation  escapes to space from to top of the cloud where it is relatively cold, whereas, in the cloudless region, more radiation to space occurs from much deeper and hotter regions. The two regions will, therefore, experience extremely different vertical profiles of  radiative heating and cooling, which will lead to horizontal temperature differences on isobars. These horizontal temperature contrasts will drive an overturning circulation that, in turn, can advect cloud condensate vertically, and in principle might be capable of maintaining the cloud patchiness.

There are two distinct cloud radiative feedbacks. The first one involves interactions between cloud formation, cloud radiative feedback and small-scale convective transport of tracers. It can result in spontaneous  variability of both the cloud and thermal structures locally in an atmospheric column that occupies a small horizontal area. This has been extensively demonstrated in \cite{tan2019} using a simple one-dimensional (1D),  time-dependent model that couples radiative transfer, cloud formation, and small-scale convective mixing.  In a large-scale sense, this generation of spontaneous variability is intrinsically 1D, without requirement of explicit 3D large-scale flows. The second feedback  with the large-scale circulation requires intrinsically multi-dimensional flows. When clouds are advected by large-scale motions, and the large-scale motions are driven by the radiative heating or cooling associated with cloud properties, such a system could be linearly unstable, providing an energy source to drive the circulation \citep{gierasch1973}.

The latter multi-dimensional cloud radiative feedback has never been investigated for giant planets in the fully nonlinear regime. There is a strong motivation to examine this cloud feedback in a full numerical model and explore its dynamical properties. In reality, this form of feedback cloud be important in atmospheres of many BDs and directly imaged EGPs, for example, for some L dwarfs or late T dwarfs in which clouds condense in the upper stratified atmospheres where convection  does not directly provide mixing (e.g., \citealp{ tsuji2002, burrows2006, morley2012}). In this situation, the intrinsic 1D variability driven by cloud feedback would not occur.  The evolution of clouds and temperature anomalies would rely on the explicit large-scale flows. 

In this study, we numerically investigate effects of the cloud radiative feedback on driving a large-scale circulation in the context of BDs and directly imaged EGPs using idealized general circulation models that couple cloud formation and their radiative feedbacks to the flows. We tune the model parameters such that clouds form in the stratified layers and the 1D intrinsic variability would not occur, leaving us a clean environment to understand the multi-dimensional cloud radiative feedback. {\btt We emphasis that this study does not directly aim at addressing the role of dynamics on the L/T transition, but to illustrate the fundamental dynamical properties of cloudy atmospheres relevant to those of dusty L dwarfs.  }

We have three basic conclusions from this study: 1) vigorous atmospheric circulation can be triggered and self-sustained by the cloud radiative feedback, providing a mechanism for surface inhomogeneous and its short-term evolution for BDs and directly imaged EGPs; 2) typical horizontal lengthscales of storms are closely related to the Rossby deformation radius, which is {\btt inversely proportional to the local rotational rate } when other parameters are the same; 3) the vertical extent of clouds decreases with increasing rotation rate.   

This paper is organized as follows. Section \ref{ch.model} introduces the numerical model. Section \ref{ch.fplane} describes results from models with varying Coriolis parameters $f$.   In section \ref{ch.discussion}  we discuss our results and implications, then finally draw conclusions in section \ref{ch.conclusion}.

\section{General Circulation Model}
\label{ch.model}

We solve the standard 3D hydrostatic primitive equations in pressure coordinates. These  are standard dynamical equations  used in dynamical meteorology  for a stratified atmosphere with horizontal lengthscales that greatly exceed vertical lengthscales (for a review, see \citealp{vallis2006}), as appropriate to the global-scale atmospheric flows in photospheres of BDs and giant planets. Two tracer equations representing condensible vapor and clouds are simultaneously integrated.  The horizontal momentum, hydrostatic equilibrium, continuity, thermodynamic, and tracer equations governing condensible vapor and  clouds are as follows, respectively,
\begin{equation}
\frac{d\mathbf{v}}{dt} = - f\hat{k} \times \mathbf{v} - \nabla_p \Phi + \mathcal{R}_{\mathbf{v}} , 
\label{eq.momentum}
\end{equation}
\begin{equation}
\frac{\partial \Phi}{\partial p} = -\frac{1}{\rho}, 
\label{eq.hydro}
\end{equation}
\begin{equation}
\nabla_p \cdot \mathbf{v} + \frac{\partial\omega}{\partial p} = 0,
\label{eq.cont}
\end{equation}
\begin{equation}
\frac{d\theta}{dt} = \frac{g\theta}{c_p T}\frac{\partial F}{\partial p}   + \mathcal{R}_{\theta} ,
\label{eq.thermo}
\end{equation}
\begin{equation}
\frac{dq_v}{dt} = -\delta\frac{q_v-q_s}{\tau_{\rm{c}}} + (1-\delta) \frac{\min(q_s - q_v, q_c)}{\tau_c}+Q_{\rm{deep}},
\label{eq.tracergas}
\end{equation}
\begin{equation}
\frac{dq_c}{dt} = \delta\frac{q_v-q_s}{\tau_{\rm{c}}} - (1-\delta) \frac{\min(q_s - q_v, q_c)}{\tau_c}- \frac{\partial (q_c V_s)}{\partial p},
\label{eq.tracercloud}
\end{equation}
where $\mathbf{v}$ is the horizontal velocity vector on isobars, $\omega=dp/dt$ is the vertical velocity in pressure coordinates, $f$ is the Coriolis parameter, $\Phi$ is the geopotential, $\hat{k}$ is the local unit vector in the vertical direction, $\rho$ is the gas density, $\nabla_p$ is the horizontal gradient in pressure coordinate, $d/dt=\partial/\partial t + \mathbf{v}\cdot\nabla_p + \omega\partial/\partial p$ is the material derivative, $\theta = T( p_0/p)^{R/c_p}$ is the potential temperature,   $p_0 = 1$ bar is a reference pressure, $R$ is the specific gas constant and $c_p$ is the specific heat at constant pressure. The ideal gas law $p=\rho R T$  is used for the equation of state for the atmosphere.   

{\btt $\mathcal{R}_{\mathbf{v}}$ represents  a Rayleigh frictional drag  applied to horizontal winds in the deep layers of our models to crudely represent the effects of momentum mixing between the weather layer and the quiescent interior  (which is  deeper than our simulation domain) where large-scale flows are likely to be retarded due to significant magnetohydrodynamic dissipation. Given the rapid convective mixing of specific entropy in the convective region and the fast rotation of BDs, the Taylor–Proudman effect may be efficient in slowing down large-scale winds in the shallow layer (see the detailed argument in \citealp{schneider2009}). Because of the much hotter interior of L and T dwarfs and likely strong magnetic fields, the conducting region in BDs is expected to extend to a  much shallower depth than that of Jupiter. Therefore this ``drag" effect is applied uniformly in our study. } The drag linearly decreases with decreasing pressure:
$\mathcal{R}_{\mathbf{v}} = - k_v(p) \mathbf{v}$,   
where $k_v(p)$ is a pressure-dependent drag coefficient, which decreases from $1/\tau_{\rm{drag}}$ (where $\tau_{\rm{drag}}$ is a characteristic drag timescale) at the  bottom pressure boundary $p_{\rm{bot}}$ to zero at certain pressure $p_{\rm{drag, top}}$:
\begin{equation}
k_v(p) = \frac{1}{\tau_{\rm{drag}}} \max \left ( 0, \frac{p-p_{\rm{drag, top}}}{p_{\rm{bot}} - p_{\rm{drag, top}}} \right ).
\end{equation}
{\btt Note that this form of drag has been used in previous studies of hot Jupiters (e.g., \citealp{liu2013,tan2019uhj,carone2020}). The strength of this drag could influence the qualitative dynamics in the cloud-forming region as will be shown in Section \ref{ch.bottomdrag}.  } In all simulations, we fix $p_{\rm{drag, top}}$ to 5 bars, which is much deeper than cloud forming regions and does not directly affect cloud formation. Kinetic energy dissipated by the frictional drag is converted to thermal heating by the term $\mathcal{R}_{\theta}$. Due to the unknown nature of interactions between the interior and the weather layer in BDs and  EGPs, the characteristic drag timescale $\tau_{\rm{drag}}$ is treated as a free parameter to explore possible circulation patterns. For major sets of simulations presented below, we adopt $\tdrag=10^5$ s.  We assume that the temperature at the model bottom boundary remains fixed during evolution, mimicking an atmosphere attached to a convective interior with a specific entropy that does not evolve over short timescales.

{\btt The deep layers of our models reach the convectively unstable region.  We parameterize effects of rapid convective mixing using a simple convective adjustment scheme as in the NCAR Community Atmosphere Model (\citealp{cam3}, see their Section 4.6). If any adjacent two layers within a vertical atmospheric column are unstable, they are instantaneously adjusted to a convectively neutral state while conserving total sensible heat $\sum \Delta p T$, where  $\Delta p$ is the layer thickness in pressure. In a single dynamical step, the whole column is repeatedly scanned until convective instability is eliminated  everywhere. Tracers are also well homogenized within the adjusted domain during adjustment. There is no adjustment in the horizontal direction. }

As in \cite{tan2019}, in calculating the net thermal radiative flux $F$ we assume a grey atmosphere (with a single broad thermal band) in a plane-parallel, two-stream, multiple-scattering approximation. The radiative transfer equations in an absorbing, emitting and multiple scattering media with the $\delta$-function adjustment for scattering are solved using the efficient numeral package TWOSTR (\citealp{kst1995}, their equations (7) and (8) are solved in our model.). The background model atmosphere uses a frequency-averaged gas opacity, the Rosseland-mean opacity $\kappa_{\rm{R, g}}$,  from \cite{freedman2014} with solar composition. The Rosseland-mean opacity gives a good estimation of radiative fluxes in the optically thick limit. In this study, clouds can extend to the upper atmosphere where  it is optically thin by the gas opacity alone. In these regions there is no good choice {\it a priori} for a single opacity in the grey approximation. Therefore, we impose a constant  minimal opacity $\kappa_{\rm{min}}$ in the whole atmosphere, such that the gaseous opacity is $\kappa_{\rm{gas}} = \max(\kappa_{\rm{R, g}}, \kappa_{\rm{min}})$. In most of the cases we chose $\kappa_{\rm{min}}=10^{-3} \;\rm{m^2~kg^{-1}}$.\footnote{We have tested additional models with  $\kappa_{\rm{min}}=0$,  $\kappa_{\rm{min}}=10^{-4}$, and  $\kappa_{\rm{min}}=10^{-2} \; \rm{m^2~kg^{-1}}$. The resulting typical circulation pattern, mean temperature structure and outgoing thermal flux  are qualitatively very similar among cases with $\kappa_{\rm{min}}=0$, $\kappa_{\rm{min}}=10^{-4} $ and $\kappa_{\rm{min}}=10^{-3} \;\rm{m^2~kg^{-1}}$. This is expected because when $\kappa_{\rm{min}}\lesssim 10^{-3} \;\rm{m^2~kg^{-1}}$, the overall thermal structure is determined by the combination of the cloud opacity and the Rosseland-mean gas opacity.  Cloud opacity usually dominates over the gas opacity, which is why clouds can drive circulation.  The case with  $\kappa_{\rm{min}}=10^{-2} \; \rm{m^2~kg^{-1}}$ exhibits a slightly different mean thermal structure because $\kappa_{\rm{min}}$ is large enough to affect the overall temperature-pressure profile even without the cloud opacity. We concluded that as long as $\kappa_{\rm{min}}\ll 10^{-2} \; \rm{m^2~kg^{-1}}$ our results are not sensitive to the choice of $\kappa_{\rm{min}}$.}   This value has been used previously for the thermal opacity of typical hot Jupiter's atmospheres \citep{guillot2010} whose atmospheric temperatures are close to those of BDs. The total atmospheric opacity $\kappa$ is simply the sum of the gas and cloud opacities $\kappa = \kappa_{\rm{gas}} + \kappa_c$, the latter being determined by instantaneous  cloud mixing ratio as a function of time and location as will be described below.

$q_v$ is the mass mixing ratio of condensible vapor relative to the dry background air and $q_c$ is the mass mixing ratio of cloud particles.  The terms $\delta(q_v-q_s)/\tau_{\rm{c}}$ and $(1-\delta) \min(q_s - q_v, q_c)/\tau_c$ are sources/sinks due to condensation and evaporation, respectively. $q_s$ is the local saturation vapor mixing ratio and $\tau_{\rm{c}}$ is a conversion timescale between vapor and condensates. We set $\delta = 1$ when vapor is supersaturated and $\delta = 0$ otherwise. The conversion timescale $\tau_c$ is very short compared to dynamical timescales in conditions relevant to BDs and EGPs \citep{helling2014}, and here we set  $\tau_c=10^2$ s which is slightly longer than a dynamical time step for numerical stability. Cloud and dust formation in substellar atmospheres is highly complex (see reviews by, for example, \citealp{helling2013,helling2019}), and many models adopt the idealized, chemical equilibrium framework (see summaries in \citealp{ackerman2001, helling2008b,marley2015}). Here we adopt an even more simplified approach---the saturation vapor mixing ratio $q_s$ is assumed to depend on pressure alone.  In this way, dynamical cloud properties (horizontal lengthscale and vertical extent of clouds) depend on dynamics alone (more specifically, rotation).  We specify a uniform pressure $p_{\rm{cond}}$, lower than which  $q_s$ decreases with decreasing pressure and higher than which $q_s$ is  arbitrarily large  such that no condensation would occur: 
\begin{equation}
\left\{ 
\begin{array}{lr}
q_s = q_{\rm{deep}} (p/p_{\rm{cond}})^3 \quad & (p\le p_{\rm{cond}}), \\
 \\
q_s = {\infty} \quad & (p > p_{\rm{cond}}),
 \end{array}
 \right.
 \label{eq.qs}
 \end{equation}
where $q_{\rm{deep}}$ is the deep vapor mixing ratio. In this study $p_{\rm{cond}}$ is assumed to be 0.5 bar, and $q_{\rm{deep}}$ is assumed to be equal to $2\times10^{-4} ~ \rm{kg/kg}$. Both choices are turned to satisfy conditions that convection is absent in the cloud forming regions while maintaining a vigorous circulation. The power law exponent of 3 in Equation (\ref{eq.qs}) is chosen non-rigorously: as long as $q_s$ sharply decrease with decreasing pressure, the resulting cloud formation is representative enough for the purpose of investigating dynamics. We have tested additional models with power law exponents of  5 and 7, and they show qualitatively similar results.     The term $Q_{\rm{deep}}=-(q_v - q_{\rm{deep}})/\tau_{\rm{deep}}$  represents replenishment of condensible vapor by deep convection over a characteristic timescale $\tau_{\rm{deep}}$, which is applied only at pressure  deeper than 5 bars. $\tau_{\rm{deep}}$ is generally set to $10^3$ s, broadly consistent with mixing timescales over a pressure scale height near the upper convective zone \citep{showman&kaspi2013}. 

We assume a constant cloud particle number  per dry air mass  $\mathcal{N}_c$ (in unit of $\rm{kg^{-1}}$) throughout the atmosphere, then use $\mathcal{N}_c$ to determine local cloud particle size with given time- and location-dependent amount of condensate $q_c$.   Cloud particles are assumed to have a single size locally in each grid cell, and the particle size $r_c$ is then determined via 
\begin{equation}
r_c = \left( \frac{3q_c}{4 \pi \mathcal{N}_c \rho_c} \right)^{1/3},
\end{equation}
where $\rho_c$ is the density of condensate.  Radiation interacts with cloud particles by absorption and scattering, which we parameterize by the extinction coefficient $Q_{\rm{ext}}$, scattering coefficient $Q_{\rm{scat}}$ and asymmetry parameter $\tilde{g}$.  As in \cite{tan2019}, the total cloud extinction opacity given a particle size $\kappa_c(r_c)$ is obtained by averaging over all wavelengths $\lambda$ using the Rosseland-mean definition:
\begin{equation}
\frac{1}{\kappa_c(r_c)} = \frac{\int_0^{\infty} \frac{1}{\kappa_{\rm{ext}}(\lambda)} \frac{dB_{\lambda}}{dT} d\lambda}{\int_0^{\infty}  \frac{dB_{\lambda}}{dT} d\lambda},
\end{equation}
where $B_{\lambda}$ is the Plank function and $\kappa_{\rm{ext}}(\lambda) =  \mathcal{N}_c \pi r_c^2 Q_{\rm{ext}}(r_c,\lambda)$ is the cloud opacity at $\lambda$.  Assuming spherical particles, the Rosseland-mean $Q_{\rm{ext}}$,  $Q_{\rm{scat}}$ and  $\tilde{g}$ as a function of temperature and particle size are calculated with  Mie theory using the numerical package developed  by   \cite{schafer2012}\footnote{https://uk.mathworks.com/matlabcentral/fileexchange/36831-matscat}.  We pre-calculate tables containing these parameters as  functions of temperature and pressure, and read them into the GCM using linear interpolation during the simulations. In this study, we use enstatite ($\rm{MgSiO_3}$) to represent properties of the cloud species, including a density $\rho_c = 3190 \;\rm{kg\;m^3}$ and the refractive index of enstatite obtained from \cite{jager2003}.  The results are not sensitive to the choice of cloud species because the essence of clouds is generating radiative heating/cooling that drives the dynamics. In all simulations we assume  $\mathcal{N}_c=10^{11} ~\rm{kg^{-1}}$, which means with the specified deep vapor abundance the condensed particle size is around 0.5 $\mu$m, consistent with the expected sub-micron particles in L dwarfs \citep{burningham2017}.  With a given particle size, the cloud settling velocity  $V_s$  as a function of pressure and temperature is calculated using  Eqs. (3)-(7) of \cite{parmentier2013}.

{\btt Our models assume a Cartesian geometry, periodic horizontal boundary conditions and a constant Coriolis parameter $f$ across the whole model domain---the so-called $f-$plane approximation.  BDs are likely rapid rotators. Doppler broadening of spectral lines  \citep{reiners2008} and rotation periods inferred from lightcurve variability (e.g., \citealp{artigau2009, radigan2012, metchev2015, apai2017,allers2020}) indicate typical rotation period of less than  two hours to slightly more than 10 hours for field BDs. Planetary-mass, free-floating giant planets and  Directly imaged EGPs likely rotate rapidly as well \citep{snellen2014, zhou2016, bryan2018}.  From equator to pole, atmospheric dynamics of BDs and directly imaged EGPs is likely affected  by a wide range of local $f$, and the key dynamical lengthscales that are closely linked to the local planetary rotation  may vary substantially.  Therefore, $f$ will be systematically varied while keeping other parameters the same in our models to thoroughly investigate effects of rotation on the circulation driven by cloud radiative feedback. }

{\btt The reason we start our investigations with a $f-$plane approximation instead of  full global models is the following. Turbulence is horizontally homogeneous under the constant$-f$ assumption, which is considerably simpler than that in the full spherical geometry in which the latitudinal-dependent $f$ induces horizontal anisotropy in the turbulence \citep{rhines1975,vallis1993}.  This strategy provides a clearer context to explore effects of varying rotation on turbulence and cloud formation. Indeed, in a long history of investigations of geophysical turbulence (e.g., \citealp{de1980,mcwilliams1999,arbic2003,arbic2007}) and Earth's tropical cyclones  (e.g., \citealp{held2008,zhou2014}), theories and models using the $f-$plane assumption have yielded significant insights on the properties of turbulence before considering a spherical geometry.   Dynamics in global models will be  investigated in a further study.}

The dynamical equations (Equations \ref{eq.momentum}-\ref{eq.tracercloud}) are solved using an atmospheric general circulation model, the MITgcm (\citealp{adcroft2004}, see also mitgcm.org). A standard fourth-order Shapiro filter is applied in the horizontal velocity and temperature fields to maintain numerical stability \citep{shapiro1970}, which smooths out grid-scale variations but has minimal effect on the large-scale structure.  We assume periodic horizontal boundary conditions. For most models presented below, we assume a  horizontal domain size of $3\times10^4$ km. The resulting area is relatively small compared to the planetary surface assuming a Jupiter sized object such that the constant $f$ approximation holds, while remaining large  enough to ensure proper statistical analysis in some cases. The pressure domain is between 10 bars and $10^{-3}$ bar, which is discretized evenly into 55 layers in log pressure. Temperature at the bottom boundary (10 bars) is fixed at 2600 K. {\btt Under cloud free conditions, the resulting effective temperature is $\sim1700$ K. With clouds, as will be shown in Section \ref{ch.fplane}, the effective temperature is $\sim1500$ K, which is appropriate for typical mid-to-late L dwarfs (e.g., \citealp{kirkpatrick2012}).} For the major models shown below, horizontal resolution is 150 km or 100 km  per grid cell depending on $f$, which  is sufficient to fully resolve dynamics within a Rossby deformation radius\footnote{{\btt The Rossby deformation radius is $R_d=c/f$  where $c$ is the phase speed of gravity wave. Using a phase speed of long vertical wave $c\sim 2NH\sim2000\mps$ where $N$ is the Brunt-Vaisala   frequency and $H$ is scale height, yielding  $R_d\sim 10^4$ to $2000$ km with $f=2\times 10^{-4}$ to $1\times 10^{-3}\;{\rm s^{-1}}$. Our horizontal resolution is sufficient to resolve dynamics within the Rossby deformation radius. }}. {\btt The dynamical time step is typically 30 s.} We adopt physical parameters relevant for BDs, including the specific heat $c_p = 1.3\times 10^4 ~\rm{Jkg^{-1}K^{-1}}$, specific gas constant $R= 3714 ~\rm{Jkg^{-1}K^{-1}}$, and surface gravity $g=1000 ~\rm{ms^{-2}}$.

{\btt Our models do not include an upper sponge layer which is typically used to minimize effects of wave reflection from the upper boundary on dynamics at levels of interest. We have performed additional tests for models with and without the sponge layer, and found that weather including the sponge layer or not do not affect the quantitative dynamical properties associated with cloud radiative feedback. These tests are shown in the supplemental document (numerical\_test.pdf) associated with this paper. In addition, we have performed a sensitivity test of horizontal resolution on a 3D model with $f=4\times10^{-4}\;\rm{s^{-1}}$. Statistical results of the test show good convergence, suggesting that our current horizontal resolution is adequate to capture the essential dynamics. The resolution test is also contained in the supplemental document (numerical\_test.pdf). We have tested additional models with different Shapiro filter timescales with either a third-order or fourth-order filter (note that the fourth-order filter which is the weaker one is used in our study).  Those results are quantitatively similar, including the mean cloud mixing ratio and kinetic energy spectrum. Therefore, the physical phenomenon seen in our results and conclusions of this study should be independent of numerical setup of the numerical diffusion. }

\section{Results}
\label{ch.fplane}

\begin{figure*}
	\includegraphics[width=2.1\columnwidth]{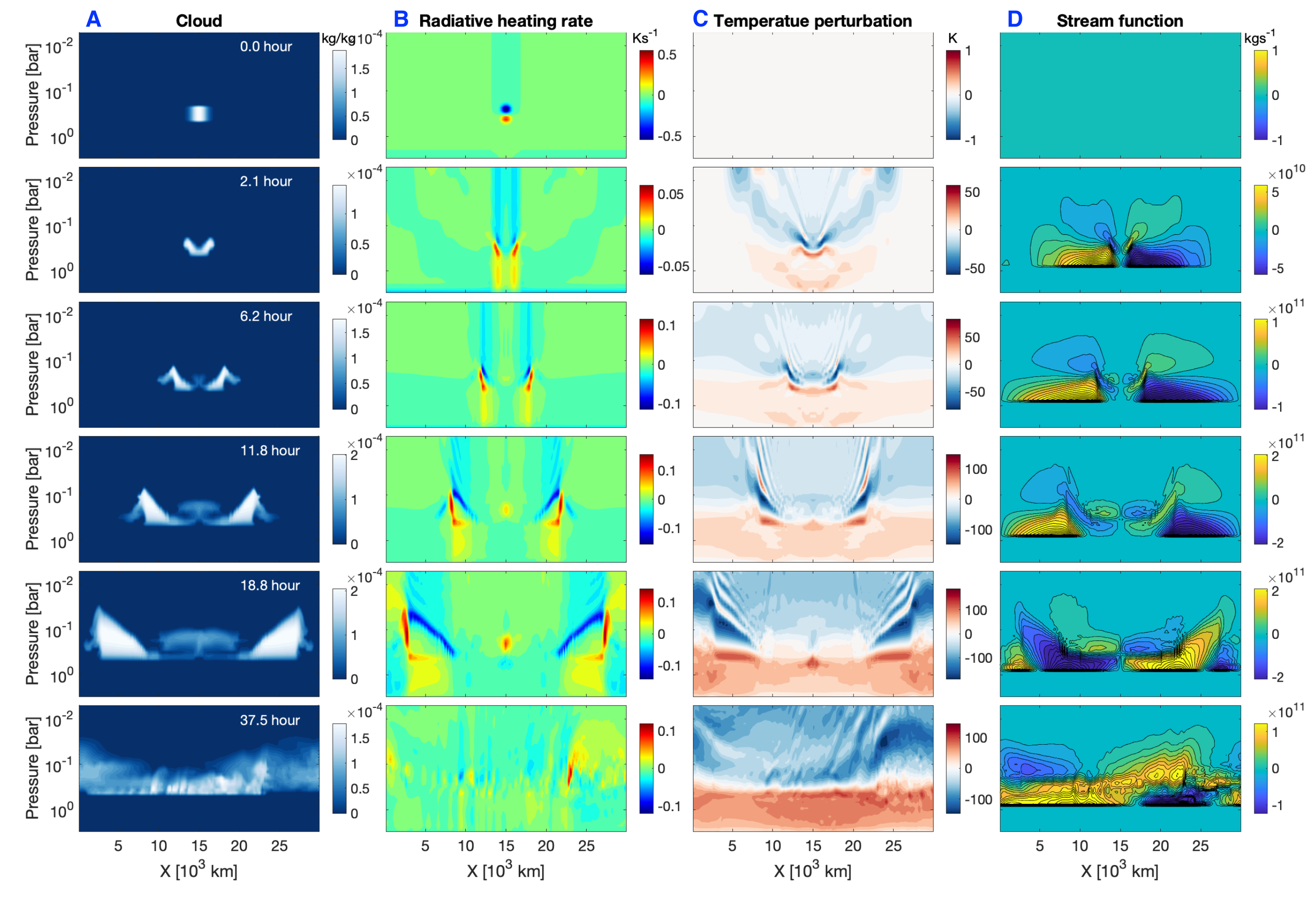}
    \caption{{\btt Time evolution of cloud distributions (column A), radiative heating and cooling rates (column B), temperature perturbations (column C) and stream function (column D) in the non-rotating 2D system. In column D, positive values represent counterclockwise and negative values are clockwise.  } Each row represents the atmospheric state at different model time indicated in the left panels. These are results of the initial evolution of a non-rotating 2D simulation.  The model is initially at rest, with an uniform initial cloud-free radiative-convective equilibrium. A patch of cloud is initially  placed  from 0.5 to 0.2 bar centered at the middle of the domain with an exponential decay in the form of $q_{\rm deep}e^{-(x'/1000 \;\rm{km})^2}$ where $x'$ is the distance from the domain center (as depicted in the upper left panel). The vapor concentration is initially $q_{\rm deep}$ and  homogeneous below 0.5 bar but is zero above 0.5 bar. The initial cloud patch generates radiative heating and cooling and drives subsequent evolution of the system.  }
\label{gravitywave}
\end{figure*}

\subsection{Non-rotating two-dimensional dynamics}
\subsubsection{Cloud radiative instability and initial evolution}
\label{ch.initial.2d}
The concept of cloud radiative instability  which was first introduced by \cite{gierasch1973}  is particularly relevant  in the cloud-driven circulation of BDs and directly imaged EGPs.   Large-scale dynamical instability may occur when the radiative heating  depends on cloud properties, and when cloud properties depend on the large-scale vertical motion  driven by the radiative heating rate.  The essence of this instability can be illustrated using a simple linearized thermodynamic system as described in \cite{gierasch1973}.    Let's assume that clouds are optically thick and that the atmospheric column was originally at rest and radiative equilibrium. The change of outgoing thermal flux could be due to brightness temperature deviations that are caused by either actual temperature variation or the cloud-top altitude variation. A slight perturbation of cloud-top altitude results  in a change in the outgoing thermal  flux. Let's suppose that the perturbation moves the cloud top upward to a colder altitude. The outgoing thermal flux decreases, and the  atmospheric column is then no longer in equilibrium and radiative heating occurs.   Large-scale vertical upwelling occurs in response to the heating, advecting the cloud top further upward (here the cloud settling speed is assumed to be smaller than the flow vertical velocity). This causes even less thermal flux emitted to space, which  induces stronger heating and ascending, providing a positive feedback to the system. 

Gierasch's theory predicts an initial growth rate $\sim \Gamma_c/(\overline{\Gamma} \tau)$ for the vertical velocity, where $\tau=\frac{c_p M}{4\sigma T^3_c}$ represents a radiative timescale, $\sigma$ is the Stefan-Boltzmann constant, $T_c$ is cloud-top temperature, $M$ is atmospheric column mass, $\Gamma_c=|dT_c/dz|$, $T_c$ is the cloud-top temperature, and $\overline{\Gamma}=\frac{d\overline{T}}{dz}+\frac{g}{c_p}$. 

\cite{gierasch1973} further showed that unstable linear modes are possible when  the cloud radiative instability is coupled to the linearized dynamical systems. In the absence of rotation, 2D hydrostatic gravity waves have a set of pure unstable growing modes (no propagation) and sets of attenuating, eastward and westward propagating modes.

The linear model of \cite{gierasch1973} serves as a valuable starting point. To better appreciate the transition from an initially linear to a  nonlinear state, we start off our numerical investigation with an initial value problem. The simplest setup is a two-dimensional  (2D) model in length-pressure domain with a periodic horizontal boundary condition  and   without planetary rotation. The system is initially at radiative-convective equilibrium in a cloud-free condition. Without clouds  it will remain motionless (notice that effect of convection is parameterized as instantaneous adjustment to a convectively neutral state and thereby has no  effect on driving large-scale  dynamics in our models). We initialize a patch  of clouds from 0.5 to 0.2 bar centered at the middle of the domain with an exponential decay in the form of $q_{\rm deep}e^{-(x'/1000 \;\rm{km})^2}$ where $x'$ is the distance from the domain center. Vapor concentration is initially set to $q_{\rm deep}$ and is homogeneous at pressures greater than  0.5 bar but is zero at pressures less than 0.5 bar. The  patch of cloud exerts substantial radiative heating at the cloud base and cooling at the cloud top, which subsequently initiate the circulation. {\btt Figure \ref{gravitywave} exhibits the time and spatial evolution of cloud mixing ratio on column A, radiative heating and cooling rates on column B, temperature perturbations relative to the initial temperature profile on column C, and stream function on column D.  The stream function of the two-dimensional model $\psi$ is defined by $u=-g/L\partial \psi/\partial p$ and $\omega=g/L\partial \psi/\partial x$, where $L$ is the horizontal domain length and $u$ is the horizontal wind. } The time sequence starts from time zero in the top row to  37.5 hours of simulated time in the bottom row.  

Interestingly, the  dynamical evolution quickly differs from that predicted by the linear theory of \cite{gierasch1973}  wherein the cloud patch was expected to initially undergo exponential growth in thickness without wave-like propagation.   Instead, radiative heating immediately drives a meridional circulation that splits the cloud patch into two parts that  propagate in opposite directions. After the splitting, the cloud layer then starts growing linearly with time while propagating  in both flanks.  After about 10 hours, a secondary, thinner cloud layer develops at the center of the pattern, which is due to the convergent flow towards the center that uplifts condensible vapor.  {\btt Eventually when the propagating clouds at different directions encounter each other (notice that the horizontal boundary condition is periodic) after around 20 hours, the model symmetry breaks and the evolution becomes chaotic. \footnote{ A movie of this initial evolution is available at \url{https://youtu.be/QZHTb20B5lY}. }} The predicted linear growth rate $\Gamma_c/(\overline{\Gamma} \tau)$ based on our modeled atmospheric conditions and a reasonable lapse rate  for clouds is on the order of $10^{-4}$ to $10^{-3}\;\rm{s^{-1}}$, greater than the growth rate of cloud thickness while they propagate horizontally (from 2.1 to 18.8 hours in Figure \ref{gravitywave}). Although the linear theory predicts a continuous spectrum of unstable modes with similar growth rates, the growing patterns in our simulation are sparse and tend to be large-scale.

The horizontal propagation of the cloud pattern is significantly slower than adiabatic free gravity waves. A crude estimation of phase speed of the adiabatic wave is $NH$ where $N$ is the Brunt-Vaisala   frequency and $H$ is scale height, yielding  $\sim1000\mps$ applying our model condition. This is much faster than the propagation of the cloud pattern, which is only roughly 250$\mps$.  A close inspection in the second row of Figure \ref{gravitywave}, especially in the panels of heating rates, temperature perturbations and stream functions, shows that a fast component already propagated more than $10\times10^3$ km horizontally away from the center in both directions at 2.1 hours.  This fast component is roughly consistent with the {\btt adiabatic free waves} that are triggered by the cloud radiative heating.  Strikingly, in later times, vertical wave-like patterns that alternate with decreasing pressure are present  above the cloud layers (see panels of heating rate and temperature perturbations at 6.2, 11.8 and 18.8 hours). These patterns do not propagate away like the fast component  at 2.1 hours, but rather have a  similar horizontal phase speed as the slowly moving cloud patterns and  relatively stationary vertical phases---analogous to  the forced, quasi-stationary waves.

The  differences between our numerical results and the linear theory of \cite{gierasch1973}  in the non-rotating 2D model perhaps partly stem from the intrinsic nonlinearity of the diabatic system---the strong cloud radiative heating depends on all modes that advect the tracer, and is non-separable in some sense. This indicates that different linear modes can be ``blended" together by the diabatic heating. This perhaps is why the growing and propagating cloud patterns in Figure \ref{gravitywave} are analogous to the triggering of free gravity waves that propagate along opposite directions---the damped but propagating modes and the growing but non-propagating modes may be blended together, giving rise to the cloud patterns that simultaneously  propagate in the horizontal directions and  grow in the vertical direction. In addition, the nonlinearity also comes in when  eddies grow sufficiently large. As seen in the middle panels in Figure \ref{gravitywave}, the strongest heating/cooling usually occurs at the edges of cloud patterns. Strong and complex local overturning circulations associated with the heating seem important in affecting the phase speed and growth rate of the pattern.  Understanding these detailed complexities is beyond of the scope of this study.

\subsubsection{At statistical equilibrium}
\label{ch.stat.2d}
The circulation does not decay away but eventually evolves towards {\btt a chaotic and self-sustained state. After about 40 simulation days, the model reaches a statistical equilibrium state  in which gravitational settling of cloud particles  is statistically balanced by net upward transport of  clouds and condensible vapor. Note that there is still time variability around the time-mean climate state after the model reaches the statistical equilibrium.} 

Available potential energy (APE) is generated from the spatially inhomogeneous radiative heating/cooling associated with partial cloud coverage, which is converted to kinetic energy (KE) associated with the vigorous circulation. The circulation in  turn maintains patchy clouds that are responsible for generation of  APE.   Kinetic energy is  removed mostly by the deep frictional drag, returning to the system via dissipated heat in the deep layers.  The fixed bottom boundary temperature (in reality  the hot interior of BDs or EGPs) ensures continuous energy supply that offsets radiation to space.   The fact that clouds must eventually settle out at low pressures is essential to maintain the circulation---otherwise without settling, clouds will be eventually homogenized above the  condensation level and therefore no patchiness would be available to generate APE. The concept of heat engines may be appropriate for the energetics of this system---it has been applied to similar top-cooling-bottom-fueling Earth moist convection \citep{renno1996} and tropical cyclone systems \citep{emanuel1986}. 

The typical isobaric  temperature perturbations in our model can exceed 100 K, and local horizontal wind  speeds can exceed 1000$\mps$. Clouds can be easily transported over 3  scale heights above the condensation level.   Winds, temperature perturbations and  clouds  exhibit a  wide range of spatial patterns, with horizontal lengthscales ranging about 1000 km to that comparable to the domain length, $ 3\times 10^4$ km. Structures with lengthscales over $2\times 10^4$ km are dominant in the vertical transport of clouds and kinetic energy, which evolve over a characteristic timescale of more than 10 hours. The seemingly dominant structure is easier seen in 2D simulations with extended horizontal  domains up to $48\times 10^4$ km shown in Appendix \ref{ch.resolution}.   The dominant  lengthscale for cloud  patches is between $2\times 10^4$ to $4\times 10^4$ km,  superposed with numerous smaller-scale structure. 

One could make use of a quasi-balance point of view to qualitatively understand why the dominant structure emerges at a horizontal lengthscale of $2\times 10^4$ to $4\times 10^4$ km.  In statistical equilibrium, large-scale subsidence through a hydrostatic, stratified atmosphere must be accompanied by diabatic cooling (an obvious example would be the Hadley circulation in Earth's troposphere, e.g., \citealp{pierrehumbert2010}).   In a presumed quasi-stationary overturning circulation, air in the cloudy regions is heated and rises, and it must subsequently cool somewhere and descend due to the requirement of continuity. The rate of radiative cooling determines the  velocity of subsiding air, which can then be used to constrain the typical horizontal lengthscale of the overturning circulation via continuity. 

Now we carry out a simple diagnostic scaling exercise based on the governing dynamical equations (Eq.[\ref{eq.momentum}] to [\ref{eq.thermo}]) to quantify the above picture. More detailed explanation of the scaling  based on a similar set of dynamical equations can be found in, e.g., \cite{komacek2016}  for application on hot Jupiter circulation. It is convenient to cast the primitive equations in log-pressure coordinates (e.g., \citealp{andrews1987,holton2012}).  We assume statistical balance (thus the time-dependent terms disappear),  and a symmetry between the ascending and the descending branch of the circulation, that they have similar speed, magnitude of heating/cooling rate and thus the fractional area. First, the continuity equation can be expressed as an order-of-magnitude estimation: 
\begin{equation}
    \frac{\mathcal{U}}{\mathcal{L}} \sim \frac{\mathcal{W}}{H},
    \label{eq.continuity.scaling}
\end{equation}
where $\mathcal{U}$ is the characteristic horizontal velocity, $\mathcal{L}$ is a characteristic lengthscale  of the circulation,  $\mathcal{W}$ is a characteristic vertical velocity with a unit$\mps$, and $H$ is scale height. In the angular momentum equation, because of  the absence of rotation, the major balance is expected between the pressure gradient and  advective forces, which in order-of-magnitude is $\mathcal{U}^2/\mathcal{L} \sim \Delta \Phi/ \mathcal{L}$ (e.g., \citealp{tan2017}). Hydrostatic balance relates the isobaric temperature gradient to the isobaric geopotential gradient. The vertical difference of geopotential $\delta \Phi$ in a single column is $\delta \Phi= -R\overline{T}\delta \ln p$,  where $\overline{T}$ is an appropriate vertically-averaged temperature  and $\delta \ln p$ is a characteristic thickness of the atmospheric column that is affected by the cloud radiative feedback in log pressure. Thus, the characteristic isobaric geopotential difference between  two columns that have a characteristic isobaric temperature difference $\Delta T$ is $\Delta \Phi \sim R\Delta T\delta \ln p$. Combining with the angular momentum equation, we have 
\begin{equation}
    \mathcal{U} \sim \sqrt{R\Delta T\delta \ln p}.
    \label{eq.am.scaling}
\end{equation}
The balance in the thermodynamic equation is expected to be between   radiative heating/cooling and advection, and is the same for both the ascending and descending branch: $\frac{\delta F}{c_p \delta M}  \sim   {\rm max} \left[ \frac{U\Delta T}{\mathcal{L}}, \frac{\mathcal{W}N^2H}{R} \right]$, where $\delta F$ is the radiative flux difference in-and-out of the atmospheric column  with mass $\delta M$. The two terms on the  right hand side are horizontal and vertical heat transport, respectively. Based on our simulated  results, the term $N^2H^2/R$ is usually larger  than $\Delta T$, so together with  the continuity Equation (\ref{eq.continuity.scaling}) we expect that the balance in  the thermodynamics  is between vertical advection and radiative heating/cooling:
\begin{equation}
    \frac{\delta F}{c_p \delta M}  \sim    \frac{\mathcal{W}N^2H}{R}.
    \label{eq.thermo.scaling}
\end{equation}

Using all  the balances in equations (\ref{eq.continuity.scaling}), (\ref{eq.am.scaling}) and (\ref{eq.thermo.scaling}), we are in the position to estimate the typical lengthscale of the overturning circulation
\begin{equation}
    \mathcal{L} \sim \frac{c_p \delta M (NH)^2}{\delta F} \sqrt{\frac{\Delta T \delta \ln p}{R}}.
\end{equation}
We apply the following numbers based on diagnostic result of our simulations: $(NH)^2\sim 10^6\mpss$, $\delta F\sim 10^5 \;\rm{Wm^{-2}}$, $\delta M=\delta p/g\sim 5\times 10^5/10^3$ kg (the thermal response of an atmospheric column to clouds extends far deeper than the assumed condensate level of 0.5 bar, and here we set $\delta p\sim 5$ bars based on our numerical results),  $\Delta  T\sim 100$ K and $\delta \ln p\sim 4$. The resulting characteristic lengthscale is about $\mathcal{L}\sim 1.9\times 10^4$ km,   consistent with numerical results.

In a short summary, in the absence of planetary rotation and under our assumed atmospheric conditions, a typical horizontal lengthscale on the order of $\sim 2\times 10^4$ km is required for the hot air to cool and return to the altitude  where it was heated up, which is a requirement for  a closed thermodynamic loop  of the air. Failing to provide a sufficiently large      simulated domain in a non-rotating 2D system may result in suppression of the self-maintained dynamics. This is because there is not sufficient room for the air to cool and descend, and thus no flow can be lifted up to form new clouds  due to continuity and the initial clouds would gradually settle down. As a result, the  response of the simulated atmosphere to the initial perturbation is simply to cool down as a whole towards the cloud-free equilibrium. Indeed, we tested a model with domain size $1.5\times 10^4$ km (half of the default value), and the model eventually became quiescent no matter how strong an initial perturbation was applied.

\begin{figure*}
	\includegraphics[width=2\columnwidth]{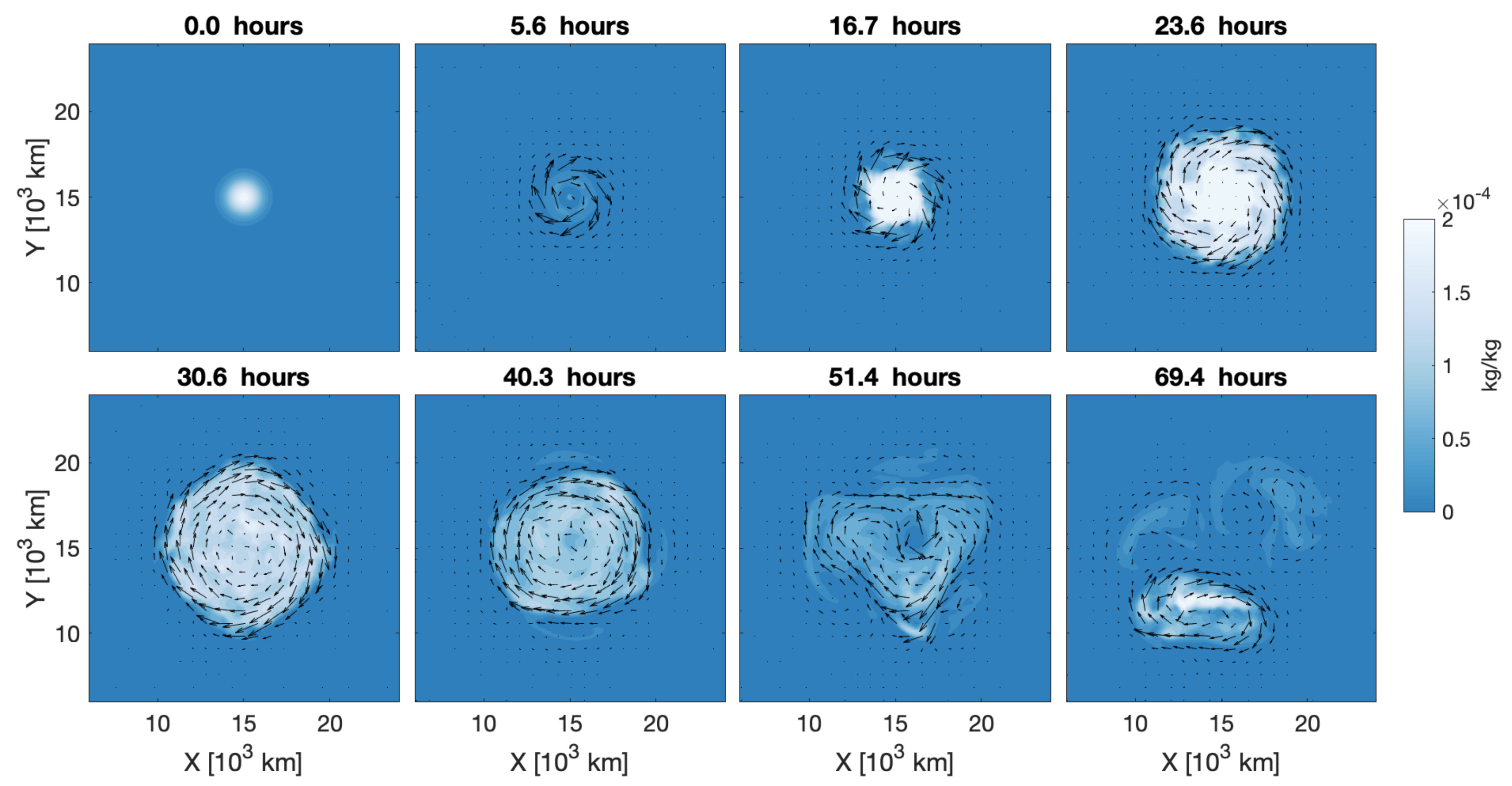}
    \caption{Time evolution of cloud distribution in a 3D  simulation with a constant Coriolis parameter $f=6\times 10^{-4} ~\rm{s^{-1}}$. Color contours in each panel represent the instantaneous cloud mixing ratio at different time and at pressure 0.23 bar, and  arrows represent the  horizontal wind vectors. The model is initially at rest, with an uniform initial cloud-free radiative-convective equilibrium, and no clouds except in a small area centered around $X,Y=15$ depicted in the upper left panel. Note that these panels are zoomed in. }
\label{fig.rotation}
\end{figure*}

\subsection{Three-dimensional dynamics and the importance of rotation}
\subsubsection{Initial evolution}
\label{ch.initial.3d}
Planetary rotation plays a fundamental role in shaping the large-scale dynamics  of rapidly rotating BDs and directly imaged EGPs \citep{showman&kaspi2013}.  One of the profound consequences of  rapid rotation on the stratified, thin atmosphere is the emergence of a major  balance between the rotation and horizontal pressure gradient, which strongly affects the scale of flows, transport of tracers and typical  horizontal temperature variations. 
Rotation leads to a natural dynamical lengthscale --- the Rossby deformation radius $R_d=c/f$  where $c$ is the phase speed of gravity wave, a lengthscale over which many types of interesting dynamical phenomenons occurs (e.g., \citealp{vallis2006}). 

In the inertia gravity wave system of \cite{gierasch1973}, when $f$ is large, unstable modes associated with cloud radiative instability are only possible for small horizontal lengthscales. For the quasi-balanced flow under rapid rotation (the so-called quasi-geostrophic flow), unstable modes are possible for axisymmetric modes, i.e., flows associated with  coherent vortices.  Growth rates of the axisymmetric modes are  greater for larger horizontal wavenumber (smaller horizontal lengthscale). The instability ceases  when the lengthscale becomes much greater than the deformation radius, because over large lengthscales the vertical motions tend to be inhibited by rotation.

Similar to the exercise  in Section \ref{ch.initial.2d}, we investigate the initial-value problem to see how rotation shapes the dynamical evolution, but now in a 3D model with a double-periodic horizontal boundary condition and a constant Coriolis parameter $f=6\times 10^{-4}\;\rm{s^{-1}}$.   Similarly, the 3D model is initialized with a patch  of clouds from 0.5 to 0.2 bar centered at the middle of the domain with an exponential decay in the form of $q_{\rm deep}e^{-(r/1000 \;\rm{km})^2}$  where $r$ is the horizontal distance to the domain center. Vapor concentration is initially $q_{\rm deep}$ and  homogeneous at pressures  larger than 0.5 bar but is zero at pressures less than 0.5 bar.  The spatial and  time evolution of cloud mixing ratio at 0.23 bar is shown in Figure \ref{fig.rotation}, in which each panel shows the instantaneous  cloud mixing ratio in color contours and the wind field in arrows.  The time  of each frame is indicated above each panel.

The initial evolution is essentially a geostrophic adjustment, a process by which an initially unbalanced perturbation is adjusted towards the geostrophic balance --- a balance between the pressure gradient force and the Coriolis force (e.g.,  \citealp{holton2012,gill1982}). Radiative heating associated with the initial  cloud patch generates a strong positive temperature anomaly over a short time  after  initialization,  which  drives a subsequent   outflow from the cloudy region.  The outward winds are then deflected towards their right  by rotation, and a strong anticyclone forms around the cloud pattern.\footnote{A cyclone has  relative vorticity the same sign as the local planetary rotation, whereas the anticyclone has the opposite sign of relative vorticity.} In this configuration, the Coriolis force tends to balance the outward pressure gradient force. Residual outward flow still persists despite the major geostrophic balance that is quickly established.  This residual  flow is responsible for the continuous horizontal growth of  the  cloud pattern until the size of the vortex saturates at about $4.5\times 10^3$ km in radius. In the classic geostrophic adjustment problem wherein an initially unbalanced field is freely evolved without forcing and dissipation, the final equilibrium height (or pressure) field is predicted to be characterized by an exponential decay with a characteristic lengthscale of a deformation radius (e.g., \citealp{kuo1997,gill1982}). The deformation radius in conditions relevant to our simulations may be estimated using a phase speed of long vertical wave $c \sim 2 NH$, yielding $R_d\sim 3.3\times 10^3$ km, which is roughly  consistent with the prediction  by the  much simpler classical geostrophic adjustment theory despite  the more complex dynamical process in our system.

The circulation and cloud pattern before about 40 hours are dominated by a basic state that is  axisymmetric  around the center of the initial cloud patch.  This is consistent with  predictions by the quasi-geostrophic cloud radiative instability \citep{gierasch1973}.  The basic state is superimposed by a seemingly wavenumber-4 non-axisymmetric component starting from 16.7 hour, and this particular  component does not amplify  as the evolution goes on. Eventually after  about 50 hours, the basic axisymmetric circulation breaks, and multiple vortices emerge from a wavenumber-3 non-axisymmetric component at the end of the sequence shown in Figure \ref{fig.rotation}.   The initial  non-axisymmetric component may be a result of  instability of the anticyclone. One possibility is that a  vortex initially embedded in an  environment at rest may be unstable due to instability analogous to  the shear instability   (e.g., see a recent work by \citealp{reinaud2019} for quasi-geostrophic vortices). 
The other possibility is the inertial instability, which may occur  when  the absolute angular momentum $\mathcal{M}=rV+\frac{1}{2}fr^2$  of the vortex decreases with increasing radial distance ($f\frac{\partial \mathcal{M}}{\partial r}<0$, e.g., \citealp{holton2012}), where $r$ is the radial distance and $V$ is the azimuthal velocity of the vortex. We have confirmed that the criteria $f\frac{\partial \mathcal{M}}{\partial r}<0$ is indeed satisfied at the pressure level shown in Figure \ref{fig.rotation} very  soon after the initialization. The unstable region is mostly near the outer edge of the anticyclone where the speed of the clockwise azimuthal wind rapidly increases with  $r$.

Compared to the non-rotating results in Figure \ref{gravitywave}, we find that rotation significantly confines the horizontal extent of the circulation  and cloud pattern.  For example, at about 16.7 hours after the initialization,   the edge  of the cloud pattern only extends out to roughly 3000 km away from the center in the case with $f=6\times 10^{-4}\;\rm{s^{-1}}$, whereas that  in the non-rotating 2D system  already reaches more than 12000 km away from the center. Our finding is qualitatively consistent with the conceptual understanding of the quasi-geostrophic, cloud radiative instability theory of \cite{gierasch1973}, showing that the typical horizontal lengthscale of the cloud-radiative-driven circulation can be limited to a scale close to the deformation radius.

\subsubsection{Dynamics at statistical equilibrium with varying $f$}

\begin{figure*}
	\includegraphics[width=1.8\columnwidth]{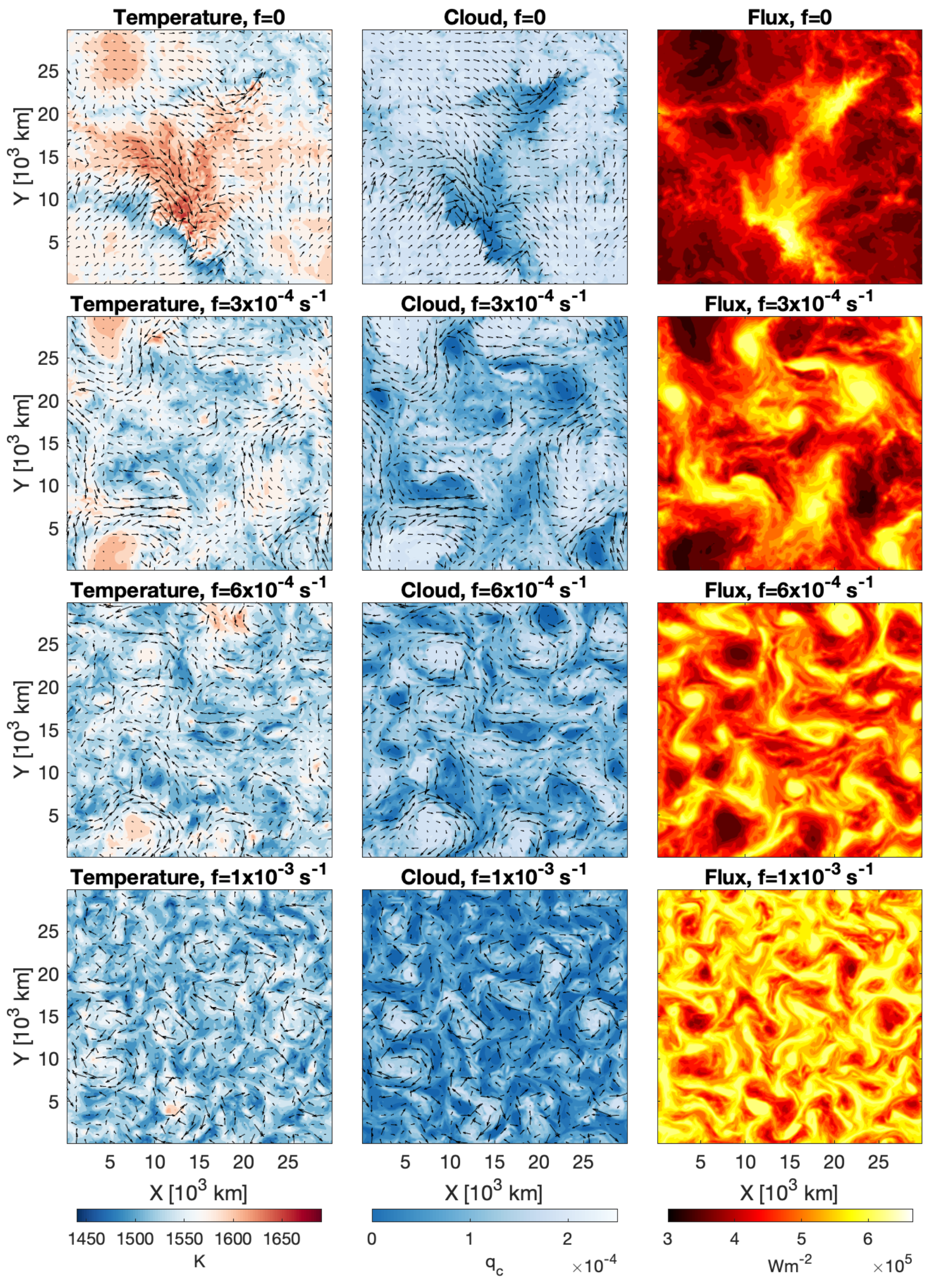}
    \caption{Instantaneous horizontal cloud mixing ratio at 0.23 bar on the left column, and the corresponding temperature at 0.23 bar on the middle, and the corresponding outgoing top-of-atmosphere thermal flux on the right. Arrows represent instantaneous horizontal wind vectors. These are results from models with different Coriolis parameters $f=0$ (top row), $3\times 10^{-4}$ (second row), $6\times 10^{-4}$ (third row) and $1\times 10^{-3}~\rm{s^{-1}}$ (bottom row). The model domain size is fixed at $30000\;\rm{km}\times 30000\; \rm{km}$. Other  parameters are the same among these models.  }
\label{fig.differentf}
\end{figure*}

Similar to the 2D system, the 3D system shown in Figure \ref{fig.rotation} eventually evolves to a highly nonlinear and chaotic state in statistical equilibrium   that is self-sustained by cloud-radiative feedback.  To systematically investigate the effect of  rotation on the equilibrated states, we first performed a set of 3D simulations with varying Coriolis parameters from $f=0$ to $1\times 10^{-3}~\rm{s^{-1}}$ over a fixed domain size ($30000\;\rm{km}\times 30000 \;\rm{km}$). Selected results of the equilibrated simulations with $f=0$, $3\times 10^{-4}$, $6\times 10^{-4}$ and $1\times 10^{-3}~\rm{s^{-1}}$ {\btt (corresponding to $f$ at polar regions of objects without rotation and  with rotation periods of 11.6, 5.8 and 3.5 hours, respectively)} are shown in Figure \ref{fig.differentf}, in which the left column shows instantaneous cloud mixing ratio as 0.23 bar with horizontal wind vectors represented by arrows, the middle column shows the corresponding instantaneous temperature field with wind vectors at 0.23 bar, and finally the corresponding  top-of-atmosphere thermal flux is shown the right column. Panels at different rows have different Coriolis parameters $f$ as indicated above each panel.

\begin{figure}
	\includegraphics[width=1\columnwidth]{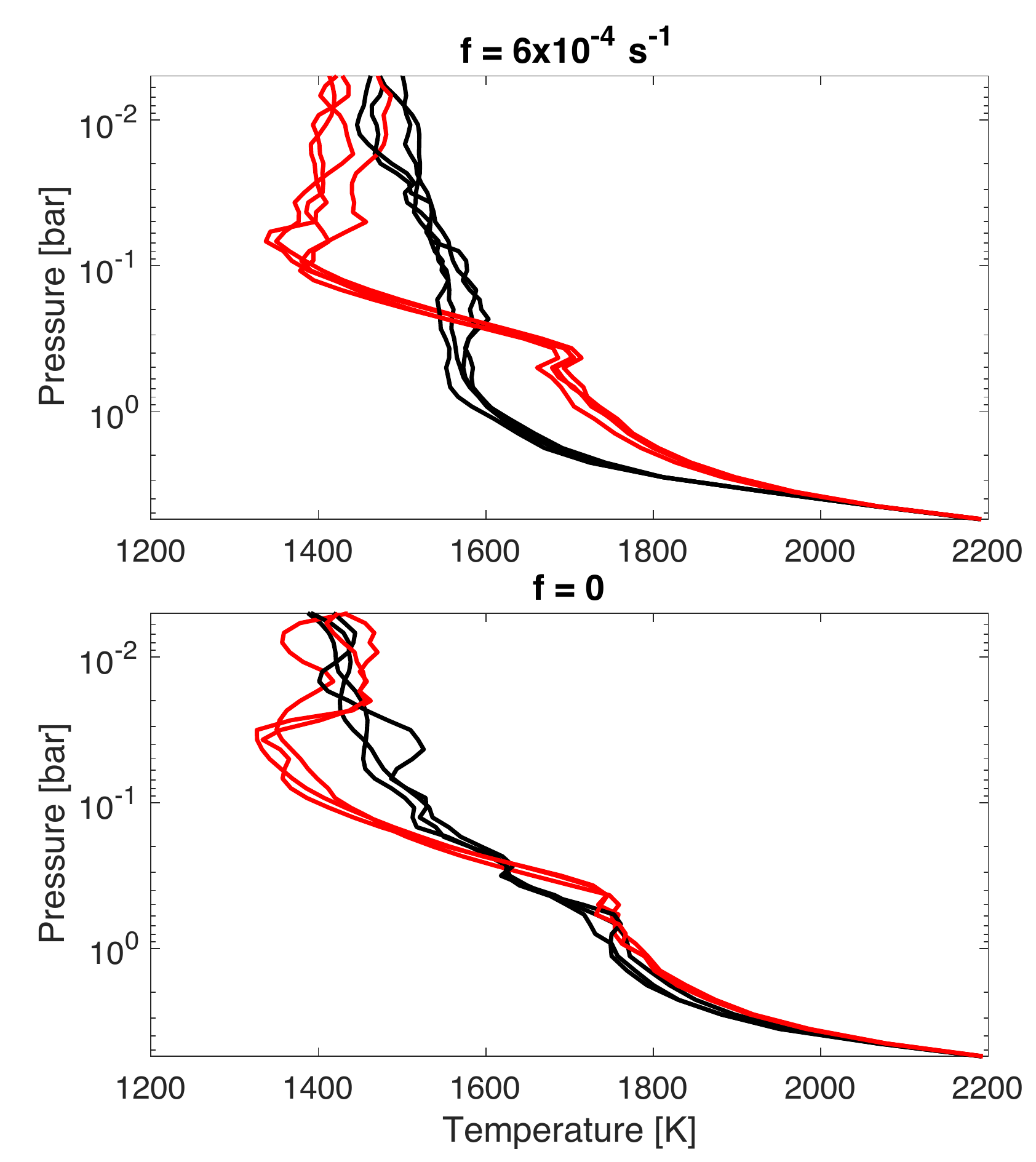}
    \caption{Selected instantaneous temperature-pressure profiles for cloudy regions (red lines) and  cloudless regions (black lines) for simulations with a Coriolis  parameter $f=6\times 10^{-4} \;\rm{s^{-1}}$ in the top panel and $f=0$ in the bottom panel. }
    \label{fig.vertical_f}
\end{figure}

The typical sizes of storms and vortices decrease with increasing rotation, as can be seen in Figure \ref{fig.differentf}. When there is no background rotation ($f=0$), dynamics in the 3D simulation are characterized by domain-scale convergent and divergent flows which are qualitatively similar to those in the non-rotating 2D simulation. When  rotation is included,  the simulated domain at 0.23 bar is populated with anticyclones that are  warmer and cloudy, and cyclones that are associated with relatively cloudless and cooler regions. Significant isobaric temperature variations over 100 K are present in all models.  There exists significant variation of the isobaric cloud mixing ratio, ranging from almost zero to more than the deep mixing ratio $q_{\rm deep}$. Cloudy  regions are usually accompanied with vigorous upwelling and cloud-free regions are associated with strong downwelling. The outgoing thermal flux exhibits strong variation across the domain, with more than $6\times10^5 \;\rm{Wm^{-2}}$ at cloudless regions and only slightly more than $3\times10^5 \;\rm{Wm^{-2}}$ at cloudy regions. This is consistent with our expectation that cloud opacity determines the level from which thermal flux escapes to space.   Interestingly, anticyclones with cloud formation almost always have larger  sizes than the cloudless cyclones.

\begin{figure*}
	\includegraphics[width=2\columnwidth]{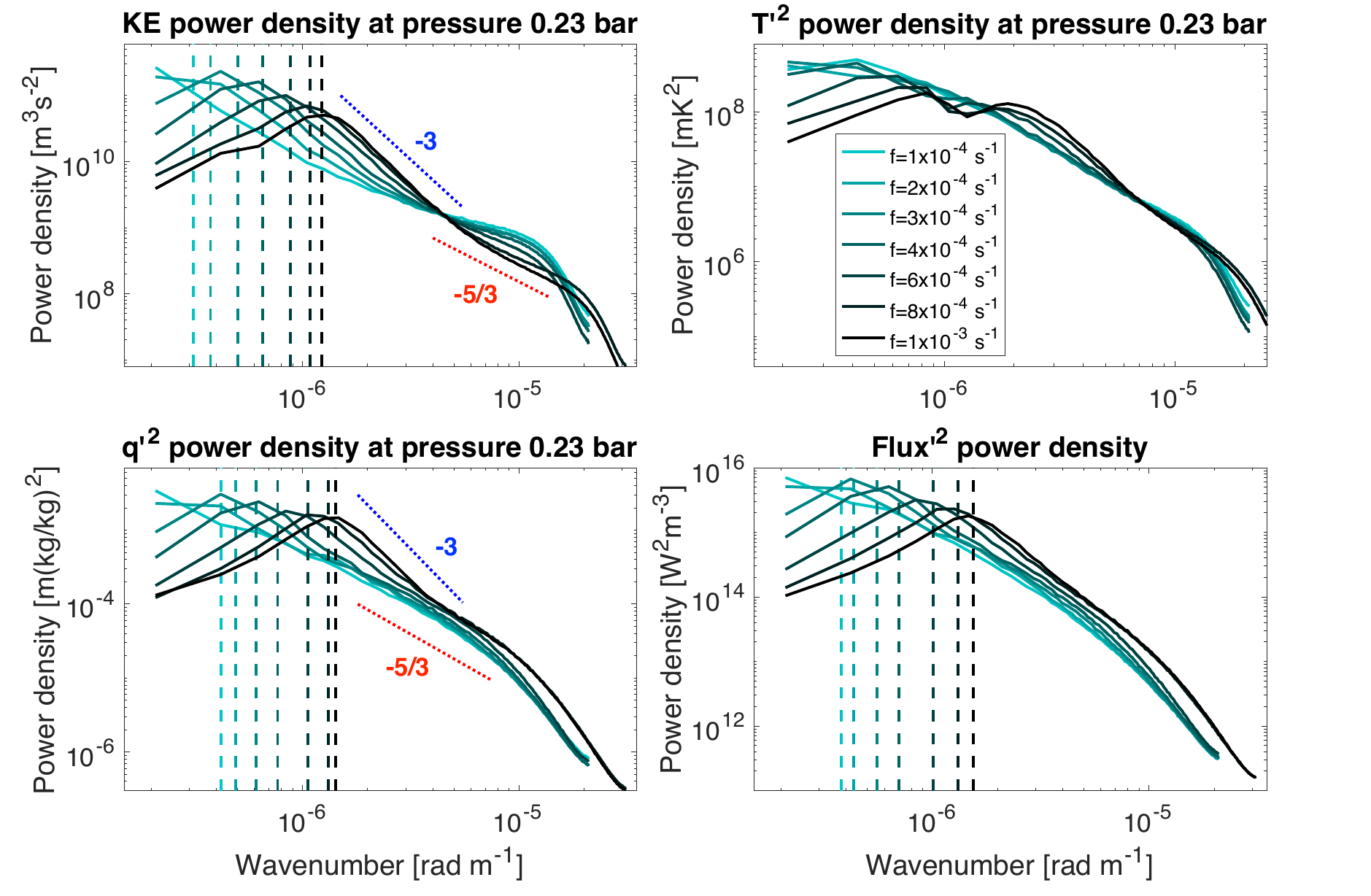}
    \caption{Time-averaged power spectra of kinetic energy (KE, top left), variance of isobaric  temperature perturbations $T'^2$ (top right), variance of isobaric total tracer perturbations $q'^2$ (bottom left) and variance of the outgoing top-of-atmosphere thermal flux $F'^2$ (bottom right) as a function of the total wavenumber $k$. The former three quantities are taken at pressure 0.23 bar. Different line colors are for cases with different Coriolis parameters from $f=1\times 10^{-4} \;\rm{s^{-1}}$ to $f=1\times 10^{-3} \;\rm{s^{-1}}$. Spectral slopes of $-3$ and $-5/3$ are indicated as dotted lines in the panels for KE and $q'^2$. Vertical dashed lines represent the energy containing wavenumber $k_e$ of each power spectra (see definition of $k_e$ in the text). $k_e$ is not plotted for $T'^2$ spectra because the presence of double peaks in some spectra  makes $k_e$ less representative. }
\label{fig.spectrum_f}
\end{figure*}

Vertical thermal profiles are shaped by both the cloud-radiative effect and the dynamics. Figure \ref{fig.vertical_f} shows several instantaneous temperature-pressure (T-P) profiles sampled in the cloudy (red lines) and cloudless (black lines) regions of the model with $f=6\times 10^{-4} \;\rm{s^{-1}}$ in the top panel and those sampled in the model with $f=0$ in the bottom panel. In the rapidly rotating case, the cloudy regions exhibit a characteristic cold top and hot bottom structure, whereas the cloudless regions are more isothermal above 1 bar which is  closer to the thermal structure that would occur in cloud-free radiative-convective equilibrium. 
In the cloudy regions, clouds form above the condensation level (which is assumed  at 0.5 bar), and their greenhouse effect warms up the air below the condensation level. The greenhouse effect extends down to about 5 bars, below which the temperature profiles merge to the same deep adiabatic profile that is specified by our bottom boundary condition. Above the condensation level, the atmospheric lapse rate $d\ln T/d\ln p$ is larger in the cloudy region  because of the top cooling and bottom heating due to cloud opacity. Clouds usually extend more than one scale height above the condensation level, which, together with the large lapse rate, result in a characteristic crossing point between two sets of T-P profiles above the convective zone. Thermal flux escapes from cloudless regions at pressures of several  bars where is hot, and thermal flux of cloudy regions is emitted from the much colder cloud-top. 

Dynamics is important to determine the thermal structure seen in Figure \ref{fig.vertical_f} both explicitly and implicitly. Explicitly, temperature fluctuations with smaller vertical wavelength are caused by the inertia gravity waves. They are present in both cloudy and cloudless regions. Implicitly, although cloud opacity is the direct cause of the typical thermal structures, both  the  cloud and temperature  anomaly  need to resist fast-travelling gravity waves that efficiently remove anomalies. Rotation plays an essential role in two key ways.  First, the quasi-geostrophic balance due to rotation helps to sustain a large isobaric temperature difference \citep{charney1963}. Second, local overturning circulations associated with vortices formed under rotation are likely important for vertical transport of clouds, and they typically evolve much slower than gravity waves. Therefore, the vertical cloud structure can  be sustained long enough for the radiative effects to be effective. Indeed, T-P profiles in the non-rotating simulations (the bottom panel of Figure \ref{fig.vertical_f}) have much smaller temperature differences between the cloudy and relatively cloudless regions.

The time evolution of the dynamical system is intriguing. Vortices are vulnerable and undergo straining, merging and dissipation over time. An individual anticyclone or cyclone usually evolves over a timescale of several to tens of hours. Inertia gravity waves are characterized by smaller length scales than the dominant vortices, evolving with a faster frequency and phase speed than the major vortices.  The thickest cloud decks are usually formed in the mature anticyclones where vigorous upwelling helps to sustain clouds against gravitational  settling.  A growing anticyclone  usually originates from a small perturbation that triggers a small patch of clouds. Then the geostrophic adjustment process discussed  in Section \ref{ch.initial.3d} is driven by the cloud radiative feedback and initialize the growth of vortices. However, not all ``seed" clouds are  able to grow. Most of the time,  small cloud patches are sheared apart or strained away by the turbulent flow, and only the lucky ones survive and are able to grow.

The size of the dominant vortices approximately  linearly decreases with increasing $f$ (Figure \ref{fig.differentf}), and now we further quantify the horizontal size distribution of various quantities.  Individual vortices are chaotic and unpredictable, and their statistical properties can be uncovered  using power spectral analysis in  wavenumber space. Assuming that the flow is statistically isotropic in the horizontal direction, we can express  the  power spectra in the total wavenumber space $|\mathbf{k}|=\sqrt{|\mathbf{k}_x|^2+|\mathbf{k}_y|^2}$ where $\mathbf{k}_x$ and $\mathbf{k}_y$ are wavenumber vector in $x$ and $y$ direction,  respectively.  Figure \ref{fig.spectrum_f} shows time-averaged power spectra for $KE =(u^2+v^2)/2$ in the upper left, variance of temperature perturbations $T'^2$ in the upper right, and variance of total tracer mixing ratio perturbations $q'^2$ at the lower left, all of which are analyzed at pressure 0.23 bar.  On the lower right we show the power spectra for the variance of outgoing thermal flux perturbations $F'^2$. These are from models with Coriolis parameters $f=1\times 10^{-4},\; 2\times 10^{-4},\; 3\times 10^{-4},\; 4\times 10^{-4}, \; 6\times 10^{-4}, \;8\times 10^{-4}$ and $1\times 10^{-3} \;\rm{s^{-1}}$. The horizontal axis is in unit of $2\pi/\rm{wavelength}$. Note that these simulations are carried out with the same simulated domain size as shown in Figure \ref{fig.differentf}.  The energy containing wavenumber, which is defined as $k_e=\left[\frac{\sum_{k\ge 1}k^{-1}E(k)}{\sum_{k\ge 1} E(k)}\right]^{-1}$, where $E(k)$ is the power at wavenumber $k$ (e.g., \citealp{schneider2009}), are plotted as vertical dashed lines for each power spectra   for $KE$, $q'^2$ and flux  variance. 

The most prominent feature of the power spectra for KE and $q'^2$ at 0.23 bar and $F'^2$ is that the peaks of the spectra systematically shift from the smallest wavenumber (the largest wavelength) to larger wavenumbers (smaller wavelength) as $f$ increases. The energy containing wavenumber $k_e$ also systematically increases with increasing $f$, although the $k_e$ sometimes differ slightly from the  wavenumber where the spectra peaks.   With a  fixed $f$, peaks of power spectra  for the above three  quantities are almost the same. For simulations with $f=1\times 10^{-4}$ and $f=2\times 10^{-4} \;\rm{s^{-1}}$, their dominant horizontal structure is comparable to the domain size and therefore their power spectra peak at the smallest wavenumber.  The power spectra of $T'^2$ at 0.23 bar differ from those for the other three quantities, especially for those with $f\ge 3\times 10^{-4}$. Not only  the spectral  peaks of $T'^2$  generally differ with those of the other three quantities given a fixed $f$, but also the $T'^2$ power spectra show double local peaks when $f$ is relatively large. The similar spectral shape  between $F'^2$ and $q'^2$ but not between $F'^2$ and $T'^2$ quantitatively demonstrates that patchy clouds (and therefore  the cloud-top temperature variations) are the dominant factors shaping the outgoing thermal flux variability,  and that isobaric temperature variation is a side effect. Indeed, analysis of  observed near-IR spectral time variability for several BDs (e.g., \citealp{buenzli2012,apai2013,lew2016,lew2020}), in which detailed static 1D atmospheric models were fit to the spectral varability, have shown that changes of cloud vertical structures are essential, and that temperature anomalies as well as the corresponding gas chemical variation help to improve the fit.

The KE spectra also help to infer the typical behaviors of the turbulence.  Pure 2D turbulence tends to transfer  kinetic energy from small to large scale starting from  where energy is injected---the so-called upscale KE transfer  of 2D turbulence.   This results in a characteristic KE power spectral slope of $-5/3$: $E(k)\propto k^{-5/3}  $ in the so-called inertial range with lengthscales larger than  the energy injection scale. In the meantime, transport of enstrophy (the square of potential vorticity integrated over the domain) from large to small scales leads to a KE power spectral slope of $-3$: $E(k)\propto k^{-3}$ in the inertial range where the lengthscale is smaller than the injection scale.  Large-scale turbulence in the rapidly rotating, stratified atmospheres has  similar properties to the 2D turbulence \citep{charney1971}. For a comprehensive tutorial of incompressible and geostrophic turbulence related to atmospheric  applications, see Chapter 8 and 9 of \cite{vallis2006}. In the KE power spectra at 0.23 bar shown in Figure \ref{fig.spectrum_f}, for  $f\gtrsim 3\times10^{-4}\;\rm{s^{-1}}$, the spectral peaks roughly locate at the scale of the internal deformation radius $L_d=c_g/f$, where $c_g$ is the phase speed of gravity waves. The KE power then decreases with increasing wavenumber in a characteristic slope close to $-3$ (a blue dotted line indicating a $-3$ slope is shown in the top left of Figure \ref{fig.spectrum_f}). However, at some point the KE spectra flattens out, and then its slope increases again at very large wavenumber.  For relatively small $f$, the spectral range with a $-3$ slope is smaller than that of  larger  $f$, and  the transition to flatter spectra occurs at a wavenumber that is closer to the peak  for cases with smaller $f$. For $f\lesssim 4\times10^{-4}\;\rm{s^{-1}}$, the KE spectral slope  can be flatter  than a $-5/3$ power law (a red dotted line indicating a $-5/3$ slope is  plotted) whereas those with $f\gtrsim 6\times10^{-4}\;\rm{s^{-1}}$ is closer to $-5/3$. Interestingly, the $q'^2$ spectra similarly exhibit a slope transition between $-3$ to $-5/3$ only when $f$ is quite large.

The spectral features suggest a few interesting dynamical processes at 0.23 bar. First, KE  at 0.23 bar is likely injected directly around the lengthscale close to the internal  deformation  radius via geostrophic adjustment. Then,  enstrophy of the turbulence cascades from the deformation radius to smaller scales, showing a characteristic KE spectral slope of  $-3$. Turbulence in the slope$=-3$ range has a quasi-geostrophic nature. Second, in the spectral space where the  slope is  flatter than $-3$, non-2D turbulence (possibly inertia gravity waves)  becomes energetically important in the KE power spectra. It has been known  that the Earth's upper troposphere exhibits a transition between a $-3$ KE spectral slope at large scale to a $-5/3$ slope at  mesoscale \citep{nastrom1984}. Non-2D turbulence likely contributes to this transition (e.g., \citealp{dewan1979,lindborg1999,lindborg2007}). In our simulations, inertia   gravity waves may play this role. Indeed, flows with scales close to the deformation radius are  close to geostrophic balance, indicating a quasi-2D nature. Flows  with much smaller scales have large  components of  imbalanced inertia gravity  waves. This was confirmed by investigating the relative fractions of the rotational and divergent parts of the horizontal velocity at different lengthscales. For models with smaller $f$, the degree of geostrophic balance is weaker, and inertia gravity waves are likely more important in the energetics than in models with larger $f$. Interestingly, the transition between slope $-3$ to $-5/3$ depends on the forcing amplitude as well. In Section  \ref{ch.forcing} in the Appendix, Figure  \ref{fig.spectrum_q},  we show that when the model  with $f=4\times10^{-4}\;\rm{s^{-1}}$ is  forced progressively weaker, the KE spectra at 0.23 bar exhibits a slope $-3$ all the way to large wavenumber before numerical dissipation takes over. When  the wavenumber continues increasing toward the largest value, numerical dissipation near the grid scale becomes important and the  spectral slope deepens.
Third, at wavenumbers smaller than that of the deformation radius (larger  lengthscales), KE is transferred upscale. However, at large scales, KE is strongly dissipated by the bottom frictional drag, so  that the KE power spectra decreases as the wavenumber decreases. We shall discuss this point more in Section \ref{ch.bottomdrag}.

\begin{figure}
	\includegraphics[width=1\columnwidth]{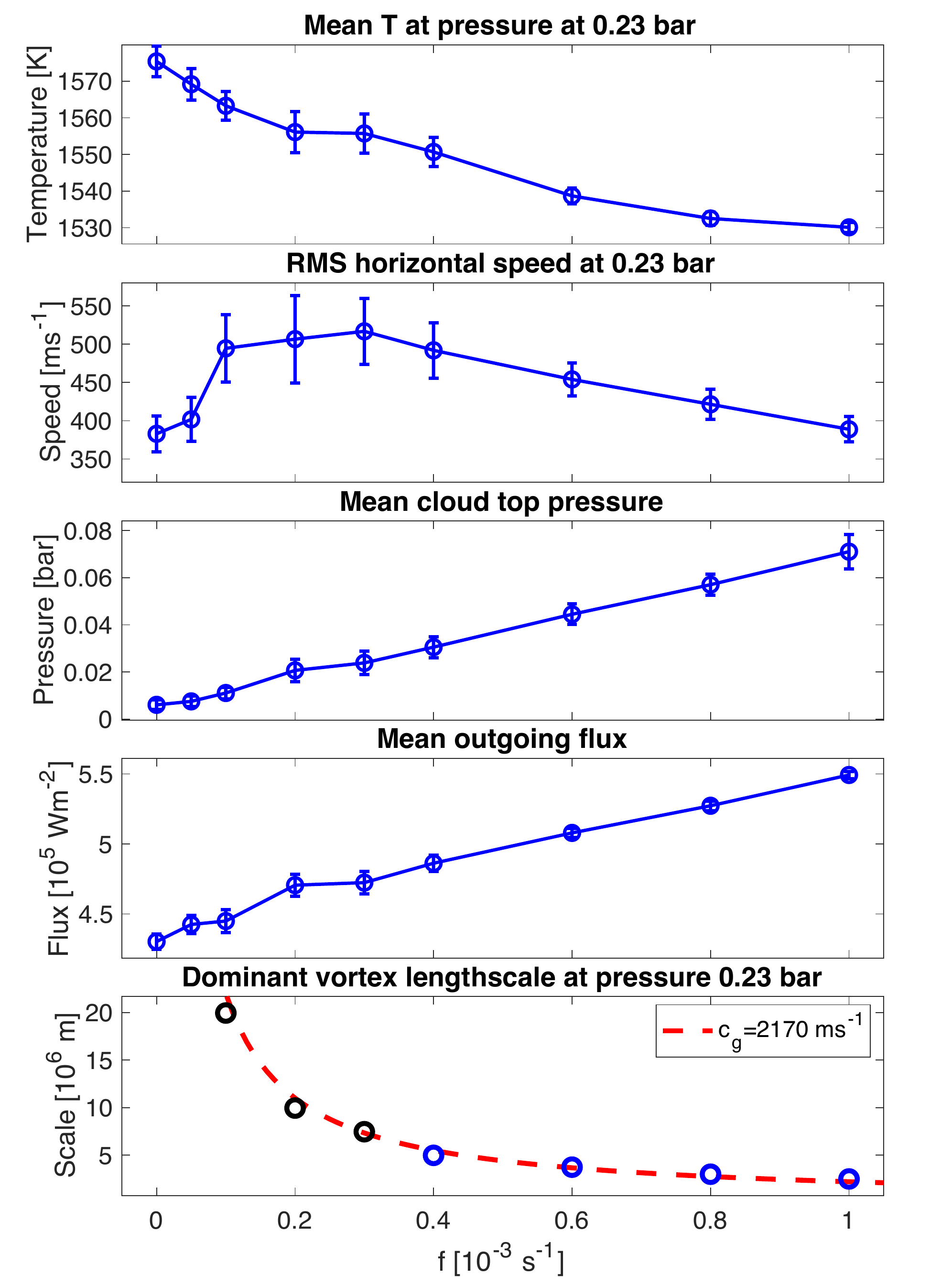}
    \caption{Various statistical quantities that are averaged over the horizontal domain and time from models with varying Coriolis parameters from $f=0$ to $f=1\times 10^{-3} \;\rm{s^{-1}}$. Errorbars in some panels are the variance of the horizontal mean quantities with respect to time, and therefore they represent the time fluctuation of the instantaneous horizontal-mean field.   All data in blue are from models with a fixed model domain size of $30000\;\rm{km}\times30000\;\rm{km}$. Panel (a) is for the horizontal temperature at 0.23 bar; panel (b) is for the RMS of horizontal wind speed at 0.23 bar; panel (c) is for the mean cloud-top pressure, which is arbitrarily defined as where the horizontal-mean cloud mixing ratio is equal to $2\times 10^{-6}\;\rm{kg\;kg^{-1}}$; panel (d) is for the outgoing top-of-atmosphere thermal flux; and finally panel (e) is for the dominant vortex size defined as $\pi/k_{\rm{max}}$, where $k_{\rm{max}}$ is the wavenumber with the maximum KE power. Note that in panel (e), black circles for $f=1\times 10^{-4}, 2\times 10^{-4}$ and $3\times 10^{-4} \;\rm{s^{-1}}$ are from models with domain size properly extended, which is to properly capture the statistical vortex behaviors. In panel (e), the dashed curve is a fit to the vortex size using the deformation radius $L_d=c_g/f$ with a fitted gravity wave phase speed $c_g$, yielding $c_g=2170 \;\rm{ms^{-1}} $. }
\label{fig.stat_changef}
\end{figure}

Finally we summarize dynamics with varying rotation in Figure \ref{fig.stat_changef} with various statistical quantities for simulations as a function of  $f$ (see the caption of Figure \ref{fig.stat_changef} for a description). There exist interesting trends with $f$. Mean temperature at 0.23 bar monotonically decreases with increasing $f$, while the mean outgoing thermal flux and mean cloud top pressure monotonically increase with increasing $f$. These three trends are  linked. As rotation increases, the mean thickness of clouds decreases, resulting in an increase of the cloud-top pressure and thus an increase of outgoing thermal flux due to the higher cloud-top temperature. The cloud  greenhouse effect is  weakened by the reduction of clouds, and thus the temperature at 0.23 bar decreases with increasing $f$. 

The wind speed at 0.23 bar shows a steep increase from $f=0$ to  $f \sim 1 \times 10^{-4}~\rm{s^{-1}}$, and then only slight increases till $f \sim 3 \times 10^{-4}~\rm{s^{-1}}$, after which it mildly declines with increasing rotation rate.  This trend probably indicates the degree of geostrophic balance established in systems with a fixed domain size. At  $f=0$, dynamics are  dominated by gravity waves. Although significant isobaric temperature differences must exist due to the cloud radiative effect, gravity waves are still efficient in removing large horizontal temperature gradients across most of the domain  (see Figure  \ref{fig.vertical_f}).  With a small $f$, a finite deformation radius emerges and vortices form, helping to sustain larger isobaric temperature variation than the non-rotating case. In this case the deformation radius  is larger or comparable to the domain size, such that slight increases of $f$ greatly help sustain a larger temperature gradient. This may help to explain the rapid trend when $f$ is small. When $f$  is sufficiently large and  the deformation radius is smaller than the domain size, more vortics  are populated in the  domain, and the trend of increasing temperature gradient flattens. When $f$ is further  increased, the deformation radius becomes a small fraction of the model domain.  In this regime, the horizontal wind speed scales as $U\sim (R\Delta T \delta \ln p)/(Lf)$ where $L$  is  a characteristic length scale of the vortices. $L$ is loosely proportional to the deformation radius and thus $Lf$ remains a rough constant. Therefore wind speed scales with the characteristic horizontal temperature variation. As has been shown, cloud radiative forcing declines with increasing $f$, and as a result, the decrease of $\Delta T$  then slows the wind speed with increasing $f$. 

As we expected, when the deformation radius is much smaller  than the domain size,  the horizontal scale of dominant vortices  linearly depends on  $1/f$ (see panel [e] of Figure \ref{fig.stat_changef}). The fit using a deformation  radius $L_d=c_g/f$ fits well to the vortex lengthscale with a gravity phase speed $c_g=2170 \;\rm{ms^{-1}} $. This is consistent with our previous estimate using a long vertical wave phase speed $2NH\sim 2000\;\rm{ms^{-1}}$.

\subsection{Vertical tracer transport}

\begin{figure}
	\includegraphics[width=1\columnwidth]{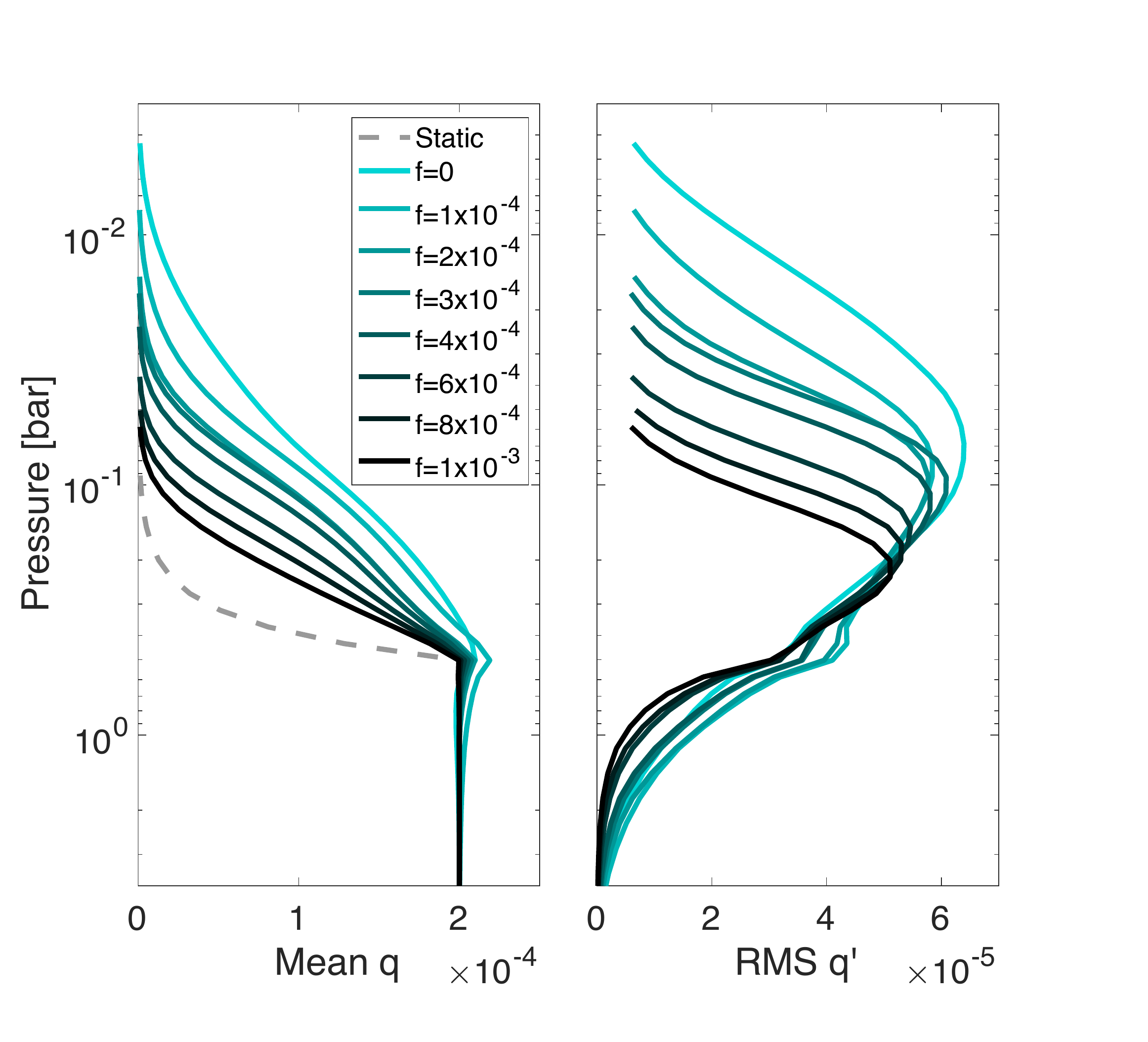}
    \caption{Left panel: the time- and horizontal-mean  mixing ratio of total tracer (gas+cloud particles) as a function of pressure for  simulations with $f=0$ to $1\times 10^{-3}~\rm{s^{-1}}$. The dashed line is the tracer profile in the absence of dynamics (i.e., it follows $q_s$ in Equation (\ref{eq.qs}) above 0.5 bar and $q_{\rm{deep}}$ below 0.5 bar).   Right panel:  the time-mean RMS of the isobaric tracer mixing ratio perturbation as a function of pressure.  }
\label{fig.patchiness}
\end{figure}

\begin{figure*}
	\includegraphics[width=2\columnwidth]{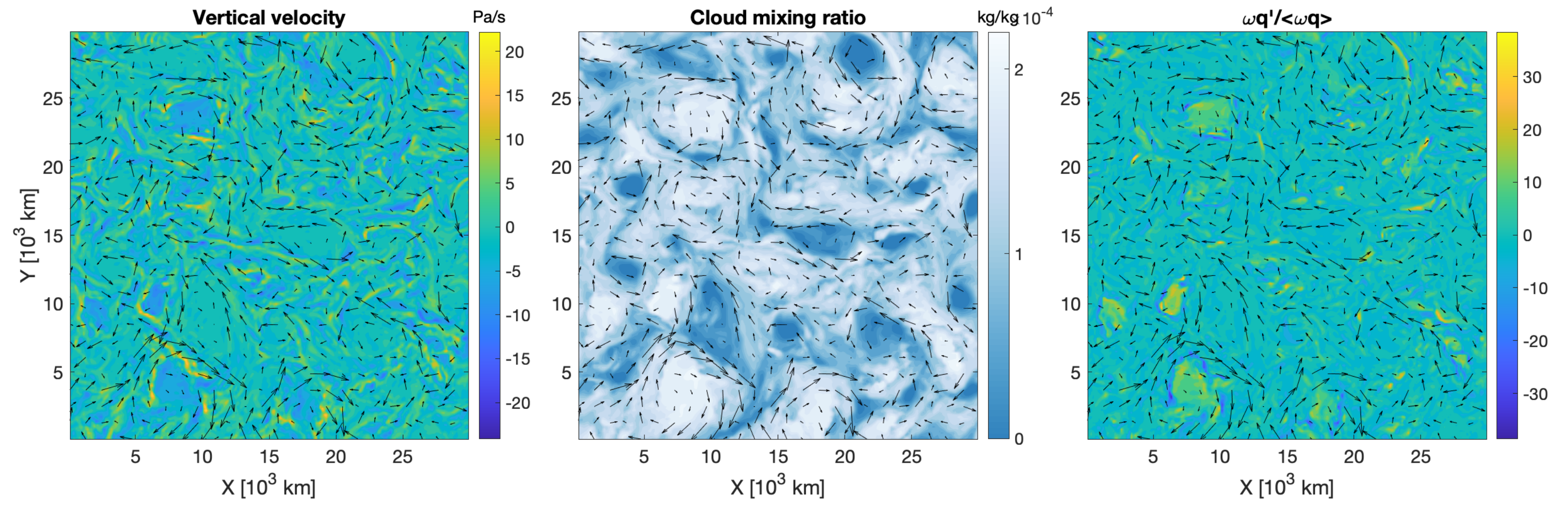}
    \caption{\btt Left panel: instantaneous vertical velocity at pressure coordinates at 0.23 bar for a simulation with  $f=6\times 10^{-4}~\rm{s^{-1}}$. Arrows represent  horizontal wind vectors.  Middle panel: instantaneous cloud mass mixing ratio at 0.23 bar. Right panel: corresponding $\omega q'_c/\langle\omega q_c\rangle$ at 0.23 bar.  }
\label{fig.verticalvelocity}
\end{figure*}

\begin{figure}
	\includegraphics[width=1\columnwidth]{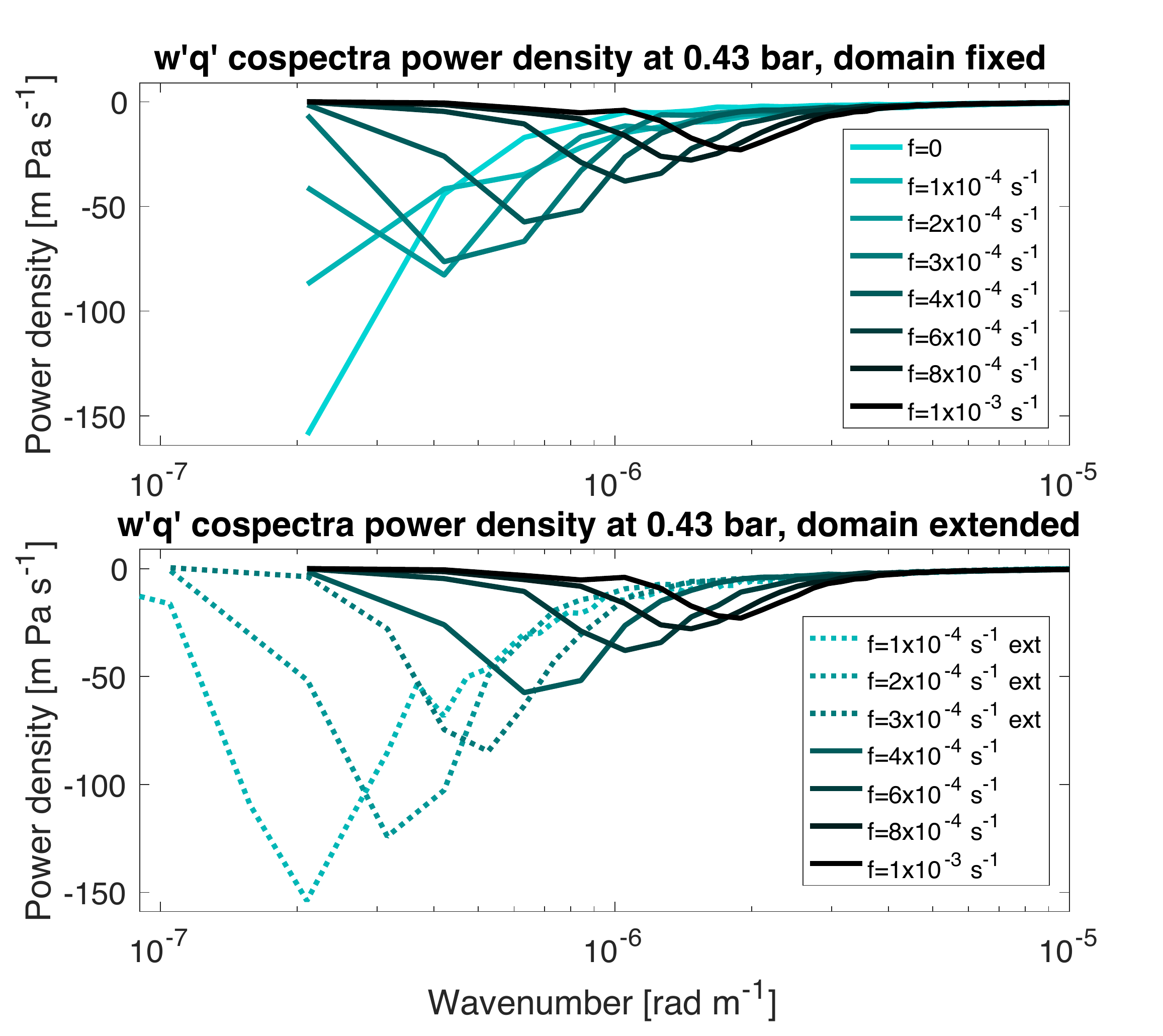}
    \caption{Top panel: cospectral power density for $\omega' q'$ at 0.43 bar (the cloud base) for models with different Coriolis parameters from $f=0$ to $1\times 10^{-3}~\rm{s^{-1}}$. All models in the top panel have a fixed domain size of $30000\times30000\;\rm{km}$. Bottom panel: dotted lines for models with  $f=1\times 10^{-4}, 2\times 10^{-4}$ and $3\times 10^{-4}\;\rm{s^{-1}}$  with properly extended domain size to  capture the statistical vortex behaviors. The same models with $f=4\times 10^{-4}, 6\times 10^{-4}, 8\times 10^{-4}$ and $1\times 10^{-3}\;\rm{s^{-1}}$ as the top panel are also plotted as solid lines in the bottom panel. }
\label{fig.wqspectra}
\end{figure}

One of the profound impacts of rotation on the dynamical system is that the thickness of cloud layers decreases with increasing rotation. We emphasize that the cloud condensation  level is  intentionally fixed at 0.5 bar at all horizontal locations, such that the change of mean cloud thickness is solely due to the change of  rotation.  The left panel in Figure \ref{fig.patchiness} shows time- and horizontal-mean total tracer (gas and clouds) mixing ratio as a function of pressure for simulations with $f=0$ to $1\times 10^{-3} \;\rm{s^{-1}}$. The right panel contains the time-mean RMS of the isobaric total tracer mixing ratio perturbation $q'$ as a function of pressure. The overall mean tracer mixing ratio  smoothly decreases with increasing rotation rate, and is the same with the RMS $q'$. Intuitively, one might imagine that the stronger the rotation, the greater the suppression   of the vertical velocity due to  the higher tendency towards geostrophic balance, making the flow less efficient to vertically transport  tracers against gravitational settling. This results in smaller horizontal temperature anomalies via the weakened cloud radiative feedback, which then give raise to a positive feedback to the reduced vertical velocity. In this subsection we quantify the net vertical transport of tracers with varying $f$ and at different pressure.

Net upward transport of tracers across an   isobaric surface relies on the positive correlation of tracer abundance and vertical velocity, i.e., having upwelling at regions where tracers are more abundant and downwelling at regions with lower tracer abundances (e.g., \citealp{holton1986,parmentier2013,zhang2018a,komacek2019vertical}). In a statistically balanced state, the horizontal mean total tracer abundance is set by a balance    between the net vertical transport by large-scale motions and gravitational settling of clouds:
\begin{equation}
    \frac{\partial (\overline{\omega'q'})}{\partial p} = - \frac{\partial (\overline{q_c V_s})}{\partial p},
    \label{eq.tracer_balance}
\end{equation}
where $q=q_c+q_v$ is the total tracer and $\overline{A}$ denotes a horizontal average of quantity $A$ over the domain. Integrating Equation (\ref{eq.tracer_balance}) from very low pressure where tracers are negligible to an arbitrary level where clouds are abundant, one obtains $\overline{\omega'q'} = - \overline{q_c V_s}$ at that level. This states that the net upward transport of total tracers across that isobar balances the total settling flux above that isobar. Regions with abundant clouds likely  have  strong radiative heating near the cloud base and  cooling near the cloud top, whereas relatively cloudless regions usually radiatively cool. This typically results in upwelling at cloudy regions and downwelling at cloud-free regions, which naturally represents a mechanism for  net upward tracer transport against gravitational settling.

\begin{figure}
	\includegraphics[width=1\columnwidth]{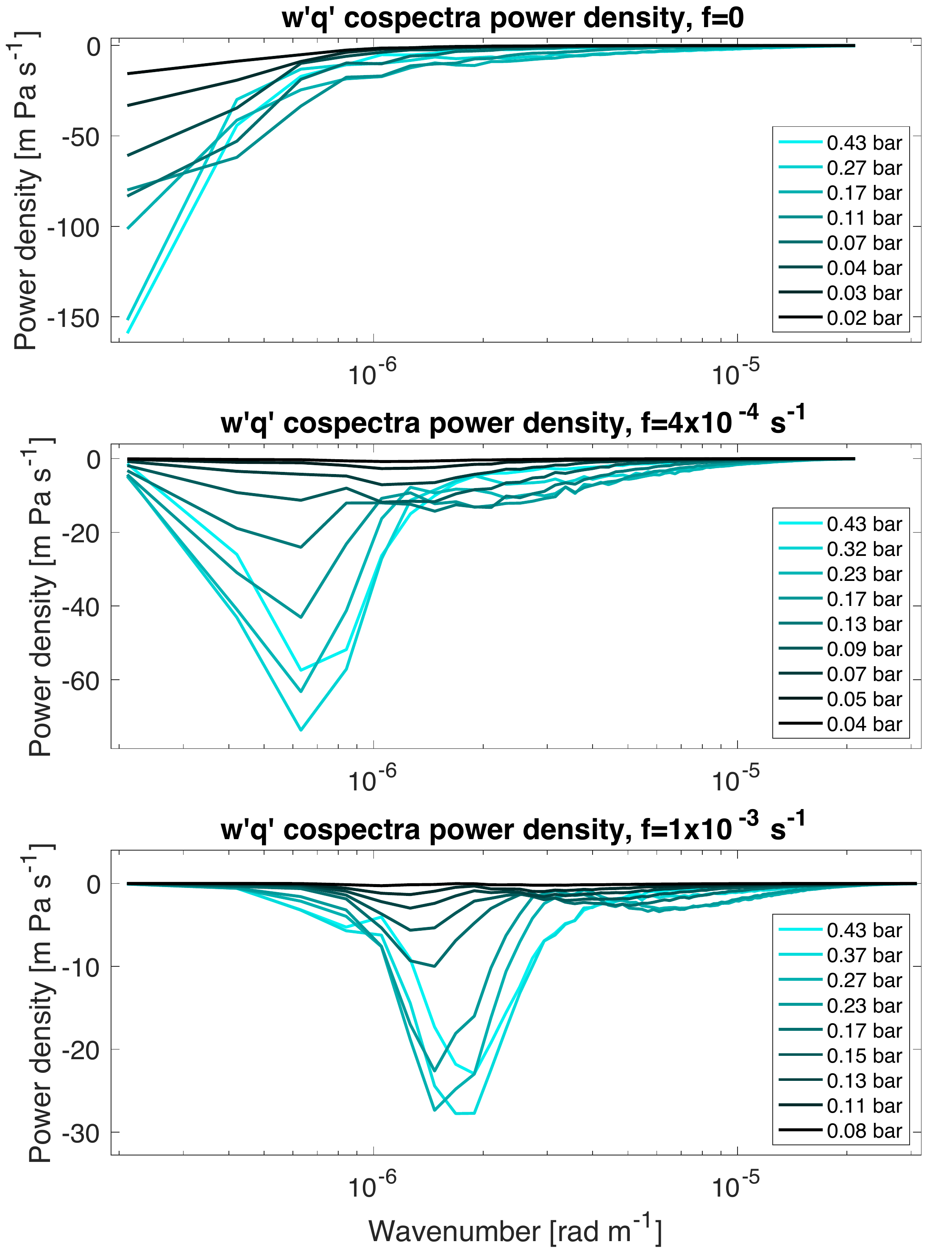}
    \caption{Cospectral power density for $\omega' q'$  at different pressures for simulations with different Coriolis parameters $f=0$ (top panel), $f=4\times 10^{-4}\;\rm{s^{-1}}$ (middle panel) and $f=1\times 10^{-3}\;\rm{s^{-1}}$ (bottom panel). }
\label{fig.wqspectra_allpressure}
\end{figure}

The questions are, what type of motions are  responsible for the major vertical transport and how do they depend on rotation? As seen from Figure \ref{fig.differentf} and their time evolution, as well as the power spectral properties of KE and $q'$ shown in Figure \ref{fig.spectrum_f}, the size of cloud patches are  comparable to the dominant vortex scales and are highly correlated to the flow pattern. Thus we would expect that  the dominant tracer transport near the condensation level is by the overturning flow associated with  cloud-forming vortices. {\btt Figure \ref{fig.verticalvelocity} shows instantaneous vertical velocity (with a unit of ${\rm Pa\;s^{-1}}$) at 0.23 bar for the model with $f=6\times 10^{-4}\;\rm{s^{-1}}$ on the left panel, the corresponding cloud mixing ratio on the middle panel and the corresponding quantity $\omega q'_c/\langle\omega q_c\rangle$ (the brackets mean horizontal averaging)  on the right panel. The latter quantity measures the correlation between tracer abundance and vertical velocity, with positive meaning upward transport and vice versa \citep{parmentier2013}. A few mature anticyclones show obvious upwelling motions inside them (notice that the negative $\omega$ represents upwelling), and these anticyclones are cloudier than surroundings. The vortex-size upwellings in anticyclones provide strong upward transport of tracers as shown in the right panel. At the meantime, smaller-scale waves and  filaments around  vortices have strong vertical motions as well, and they significantly contribute to {\it local} vertical transport.  However, these small-scale transport are in both directions which might tend to cancel out each other, and it is unclear, by just reading the maps shown in Figure \ref{fig.verticalvelocity}, how much they contribute to the {\it  total} vertical transport.  } 

{\btt To better quantify contributions of flows with different horizontal lengthscales to the total vertical tracer transport}, as a standard exercise in meteorology, we calculate the cospectral power density for the quantity $\omega ' q'$, which is simply $2\mathbb{R}(q_k \omega^{\ast}_k) $ where $q_k$ and $\omega_k$ are the  coefficients at wavenumber $k$ space for tracer and vertical velocity, and $\omega^{\ast}_m$ is the conjugate of $\omega_m$ (e.g., \citealp{randel1991}). Figure \ref{fig.wqspectra} shows the cospectral power density at a pressure of 0.43 bar (right above the condensation level), for simulations with $f=0$ to $f=1\times 10^{-3} \;\rm{s^{-1}}$ but all with a fixed domain size in the upper panel. 

When $f=0$ or the vortex scale is comparable to the simulated domain size, the divergent and convergent flow occurs mostly on the domain scale, and therefore the dominant transport power is at the lowest wavenumbers. As $f$ increases, similar to the KE spectra,  the peak of the $\omega ' q'$ cospectral power density is at lengthscales close to the deformation radius and systematically shifts to larger wavenumber (smaller lengthscale). The peak value of the cospectral power density also monotonically decreases with increasing $f$, consistent with the weakened transport of tracers shown in Figure \ref{fig.patchiness}.

The dashed lines in the lower panel of Figure \ref{fig.wqspectra} show spectra of three additional models with extended domain sizes  for $f=1\times 10^{-4}, 2\times 10^{-4}$ and $3\times 10^{-4} \;\rm{s^{-1}}$. Cases with $f\ge4\times 10^{-4} \;\rm{s^{-1}}$ and with the original domain size  are also plotted in the same panel. Models with $f\le2\times 10^{-4} \;\rm{s^{-1}}$ no longer show peaks at the largest lengthscale. Instead, the dominant transport mode is well correlated with rotation even at small $f$ when the vortices are no longer limited by the domain size.  The $\omega ' q'$ spectra  of all these models peak at lengthscales close to the deformation radius, and the peak values also monotonically decrease with increasing $f$.  

At lower pressures, however, the vertical transport by smaller-scale motions (those associated with inertia gravity waves and filaments) start to be comparable or even dominate over that by motions at lengthscales near the deformation radius. Figure \ref{fig.wqspectra_allpressure} shows the $\omega ' q'$ cospectral power density at different pressure levels for the simulation with $f=0$ in the top panel, with $f=4\times 10^{-4} \;\rm{s^{-1}}$ in the middle panel and with $f=1\times 10^{-3} \;\rm{s^{-1}}$ in the bottom panel. In the lower two panels,  at lower pressures, the cospectra starts to split into two groups, one being the motions near the deformation radius and the other being at much smaller lengthscales. The total transport power by the small-scale groups is comparable to or larger than that by the vortex-scale motions at lower pressures.

There is likely a dynamical reason for this transition. As discussed in Section \ref{ch.stat.2d},  tracer transport  by a  thermally-driven circulation requires that rising air is heated and subsiding air cools. This  works well in the non-rotating cases, which is why tracer transport is always dominated by the lowest wavenumber at all pressure levels in the case with $f=0$   (top panel of Figure \ref{fig.wqspectra_allpressure}). For large $f$, the  thermally-driven transport works well only when the air in the ascending cloudy regions is not cooler than air in the  descending cloudless regions. For example, in  the case with $f=4\times 10^{-4} \;\rm{s^{-1}}$, the temperature in cloudy regions is generally lower than that in cloudless regions when the pressure is less than 0.17 bar. In the middle panel of Figure \ref{fig.wqspectra_allpressure} with $f=4\times 10^{-4} \;\rm{s^{-1}}$, the total transport power by smaller scales starts to be comparable to that by the vortex-scale motions at 0.17 bar. At an even lower pressure, because the direct thermal-driven circulation is likely limited due to thermodynamic constraints, eddy-driven (here the eddy refers to flows with lengthscale smaller than the dominant vortex scale) circulation becomes increasingly important in the vertical tracer transport mechanism.   


\subsection{Dynamics with varying bottom drag}
\label{ch.bottomdrag}

\begin{figure*}
	\includegraphics[width=2.\columnwidth]{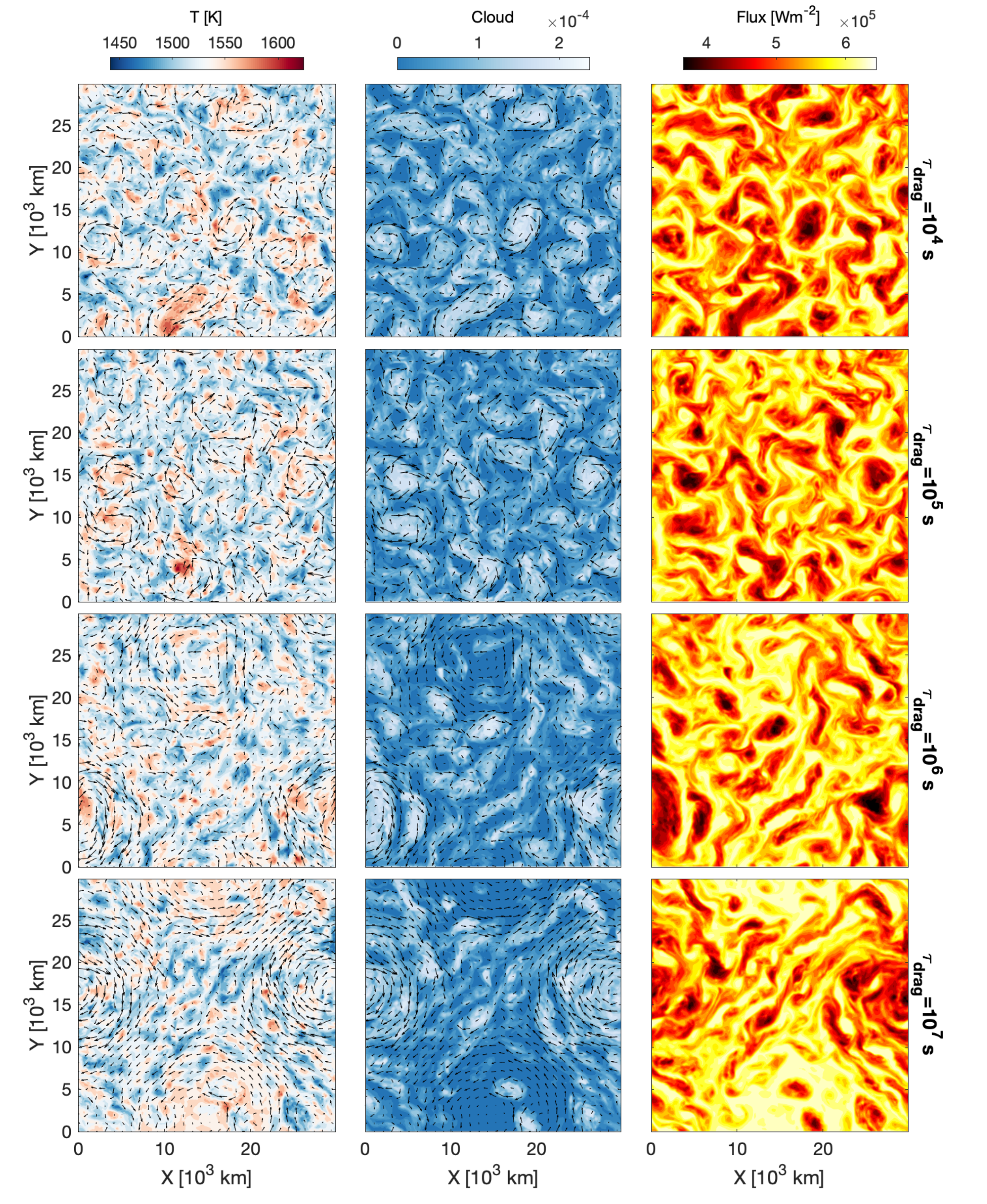}
    \caption{Instantaneous horizontal cloud mixing ratio at 0.23 bar in the left column,  the corresponding temperature at 0.23 bar in the middle column, and the corresponding outgoing top-of-atmosphere thermal flux in the right column. Arrows represent  horizontal wind vectors. These are results from models with different bottom frictional drag $\tdrag =10^4 $ s (top row), $10^5$ s (second row), $10^6$ s (third row) and $10^7$ s (bottom row). The Coriolis parameter is $f=1\times 10^{-3} \;\rm{s^{-1}}$  for all models. All other  parameters are the same among these models.  }
\label{fig.temp_drag}
\end{figure*}

Kinetic energy of the flow at lengthscales larger than the deformation radius is strongly dissipated by the frictional drag that extends from bottom up to 5 bars. Why do flows driven by the cloud formation well above 5 bars experience  bottom drag? In  quasi-geostrophic turbulence, if KE is injected from the baroclinic flow (which refers to structures with vertical variation), the upscale transfer of KE also occurs over the vertical direction in the horizontal scale close to the deformation radius (e.g., \citealp{rhines1977,salmon1978,salmon1980,smith2002,chemke2015}).  KE associated with the baroclinic flow will be transferred towards flows with greater vertical wavelength, and eventually  to the barotropic flow (which refers to flows independent of pressure or height). KE associated with the barotropic flow then continues to transfer to larger horizontal lengthscales. In our case, KE is injected by the baroclinic cloud-radiative-driven dynamics. When it is transferred towards the barotropic flow, the KE is deposited mostly in the deep layer simply because there is more mass there. With a strong bottom drag that directly removes KE of the deep layers, the rate of KE generation from the upper cloud level is not sufficient to maintain a strong barotropic flow. Therefore, in Figure \ref{fig.spectrum_f},  the strength of the domain-scale flow associated with the barotropic flows at 0.23 bar  is rather weak. 

To investigate the turbulence properties when the damping on the barotropic flow is different, we  performed additional experiments assuming different bottom drag timescales of $\tdrag=10^4, 10^6$ and $10^7$ s. This is also motivated by the fact that  the bottom drag only rather crudely represents the effect of mixing with the deep interior. There is no justification for how strong the drag should be, and here we test the sensitivity of our results to the varying  drag timescale. We assume $f=1\times 10^{-3} \;\rm{s^{-1}}$ in order to maximizes scale separation between the deformation radius and domain size.  Results are shown in Figure \ref{fig.temp_drag}, in which temperature maps at 0.23 bar are shown on the left column, maps of cloud mixing ratio are shown at 0.23 bar on the middle column, and the outgoing radiative flux is shown on the right column. The drag timescales from the top to bottom row are $10^4, 10^5, 10^6$ and $10^7$ s, respectively.  Results with $\tdrag=10^4$ and $10^5$ s appear to  be  very similar in  terms of  the typical storm  sizes, amplitudes of temperature perturbations and cloud mixing ratio. {\btt With strong drags, storms and vortices are mostly in  sizes close to the deformation radius.  When $\tdrag=10^7$ s, there appears to be two dominant horizontal modes of vortices, one being those with sizes close to the deformation radius and the other being a pair of much larger, domain-size cyclone and anticyclone. The two distinctive modes can also be easily seen  in the cloud mixing ratio and the outgoing thermal flux maps (the bottom row in Figure \ref{fig.temp_drag}). Unlike the small vortices close to the deformation radius, the domain-size vortices are dominated by a pressure-independent component.  }

\begin{figure*}
	\includegraphics[width=1.9\columnwidth]{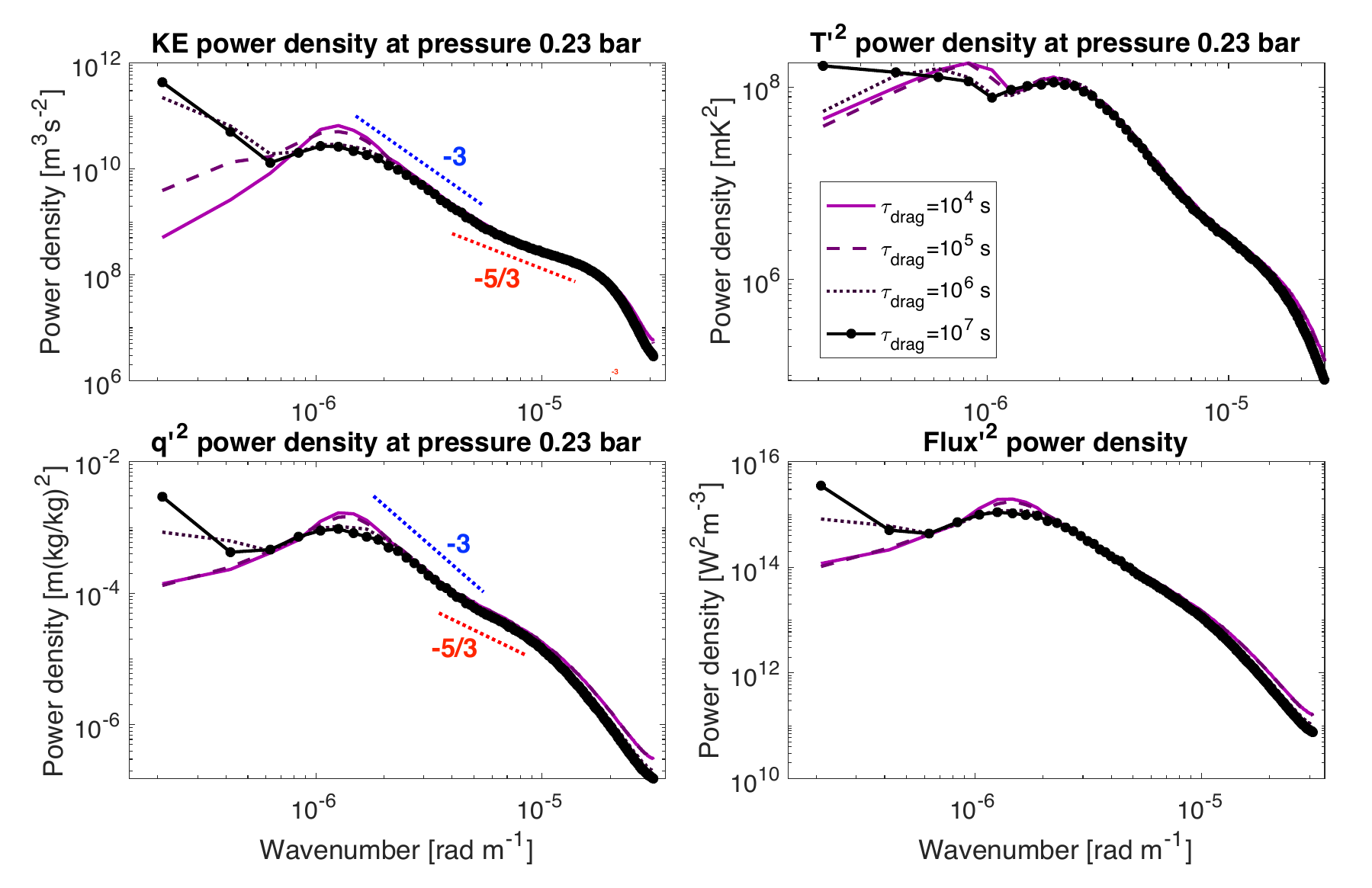}
    \caption{Similar to Figure \ref{fig.spectrum_f} but shows power spectra of KE, $T'^2$ $q'^2$ and $F'^2$ for simulations with different bottom frictional drag $\tdrag =10^4 $ s, $10^5$ s, $10^6$ s and $10^7$ s  and a fixed Coriolis parameter $f=1\times 10^{-3} \;\rm{s^{-1}}$.}
\label{fig.spectrum_drag}
\end{figure*}

We perform spectral analysis to quantify the scale separation of the  vortices. Figure \ref{fig.spectrum_drag} shows the same quantities in spectral space as in Figure \ref{fig.spectrum_f} but for the four different $\tdrag$. In the KE spectra, both  case with $\tdrag=10^4 $ and $10^5$ show a single peak at $1.2\times 10^{-6} \;\rm{rad\;m^{-1}}$, and then the KE decreases at both longer and shorter wavelengths. At smaller wavelengths (larger lengthscales), the case with $\tdrag=10^4$ s shows a lower energy density than that with $\tdrag=10^5$ s. With longer $\tdrag$, the KE spectra show two peaks, one  at $1.2\times 10^{-6} \;\rm{rad\;m^{-1}}$  and  a stronger one at the smallest wavenumber (the domain size). The KE spectra at wavenumbers larger than $3\times 10^{-6} \;\rm{rad\;m^{-1}}$ are similar among the four cases. The local peak of KE  spectral  density near  $1.2\times 10^{-6} \;\rm{rad\;m^{-1}}$ in cases with  larger $\tdrag$ is smaller than that with smaller $\tdrag$. Similarly, spectra of $q'^2$, $T'^2$ and $F'^2$ are all affected by the increasing $\tdrag$.

As the bottom frictional drag becomes weaker, the rate at which it dissipates KE decreases, and the KE associated with the barotropic flow  accumulates. Dynamics of the nearly barotropic flows are more akin to the 2D flow, and its KE in the  $f-$plane can be transferred to the domain scale if the dissipation is weak. Indeed, the simulation with  $\tdrag =10^7 $ s shows a pair of cyclone and anticyclone  comparable  to the domain size. They appear as a pair because of  angular momentum conservation. After the vortices grow to the domain size, they cease to grow in size but their KE  increases until an equilibrium between the KE upscale transfer and removal by the drag is reached. In this situation, the KE power spectrum steepens towards the largest scale.

The domain-scale vortices exert  feedbacks to the formation of storms on scales close to the deformation radius. The first prominent feature in Figure \ref{fig.temp_drag} is the existence of domain-scale isobaric temperature difference, cloud patterns and outgoing thermal flux. The domain-scale cyclone is slightly cooler, less cloudy and emits more flux to the space, whereas the anticyclone is the opposite. Unlike cloud formation with sizes near the deformation radius, these domain-size structure is not directly driven by the geostrophic adjustment.  Effects of the bottom frictional drag---analogous to  Ekman pumping and suction---acts against this configuration. For the nearly barotropic flow, the pressure gradient is mainly balanced by the Coriolis force associated with the wind in the absence of drag. In the presence of frictional forces,  the three-way force balance induces a drift in the direction from high pressure to low pressure---this means a convergent bottom flow that pushes air upward in the cyclone and divergent flow that sucks air downward in the anticyclone. This effect should in principle promote domain-scale cloud formation in the cyclone and suppress cloud formation in the anticyclone, which is the opposite to  our results.  The responsible mechanism is likely the migration of vortex under inhomogeneous background vorticity. Vortices tend to migrate to regions where the background vorticity matches the vortex's absolute vorticity. An important example  is  the migration of vortices under the planetary vorticity gradient---cyclones migrate poleward and anticyclones migrate equatorward (e.g., \citealp{adem1956,lebeau1998,scott2011}). In our model with $\tdrag =10^7 $ s, the magnitude of  relative vorticity anomalies associated with the domain-size vortices is comparable to  $f$, and therefore the background vorticity is no longer homogeneous.  Cloudless cyclones have relatively high absolute vorticity and cloudy anticyclones have relatively low absolute vorticity. When small cyclones are generated inside the  domain-size anticyclone which has low absolute vorticity, they tend to migrate to  regions with high absolute vorticity---the domain-size cyclone. Whereas cloud-forming anticyclones tend to move to the domain-size anticyclone.   Overtime, clouds tend to accumulate in the domain-size anticyclonic region and the domain-size cyclonic region is filled with cloudless air.\footnote{   We show a movie of the cloud mixing ratio at 0.32 bar (available at \url{https://youtu.be/8e4wH63bFsg}) and a movie of the potential vorticity at 0.32 bar (available at \url{https://youtu.be/TBRuh2mdWo8}) for the model with $\tdrag=10^7$ s. The relative migration between the small vortices with sizes close to the deformation radius  and the large, domain-size vortices are easily visualized in these movies.  }

Secondly, the strong shear flows associated with the strong domain-size vortices decrease the efficiency of cloud formation. Newly formed storms near the strong shear regions are sometimes disrupted and sheared apart before they become mature. This  likely causes smaller peaks around the internal deformation radius in the KE power  spectra for models with $\tdrag=10^6$ and $10^7$ s than in those with $\tdrag=10^4$ and $10^5$ s (see Figure \ref{fig.spectrum_drag}). 

Clouds induce radiative heating to the domain-scale anticyclonic region, whilst relatively cloud-free air cools the domain-scale cyclonic region. This drives upwelling in the anticyclonic region and downwelling in the cyclonic region, providing extra tracer transport in addition to the vortex migration. As a result of cloud radiative feedback, the domain-scale vortices have temperature variations on isobars.  This effect is prominent in the case with $\tdrag=10^7$ s. This is dynamically interesting because, by definition, the barotropic flow is independent of pressure (or height). Given that the bottom temperature is almost horizontally uniform in our model, a ``true" barotropic flow should not have isobaric temperature variation at any pressure. In our case, the {\btt large, domain-size vortices} is only quasi-barotropic due to the cloud radiative feedback.

{\btt In a short summary, the strength of the nearly barotropic mode is sensitive to the strength of the bottom drag. In the strong drag case, the nearly barotropic mode is efficiently damped by the drag, and dynamics in the cloud forming region is dominated by the baroclinic, cloud forming vortices with sizes close to the deformation radius. In the weak drag case, the nearly barotropic mode takes kinetic  energy from baroclinic modes and is able to grow, such that it can affect dynamics at the cloud forming levels as seen in Figure \ref{fig.temp_drag}. }

\section{Discussion}
\label{ch.discussion}

\subsection{Observational implications}
\begin{figure}
	\includegraphics[width=1.\columnwidth]{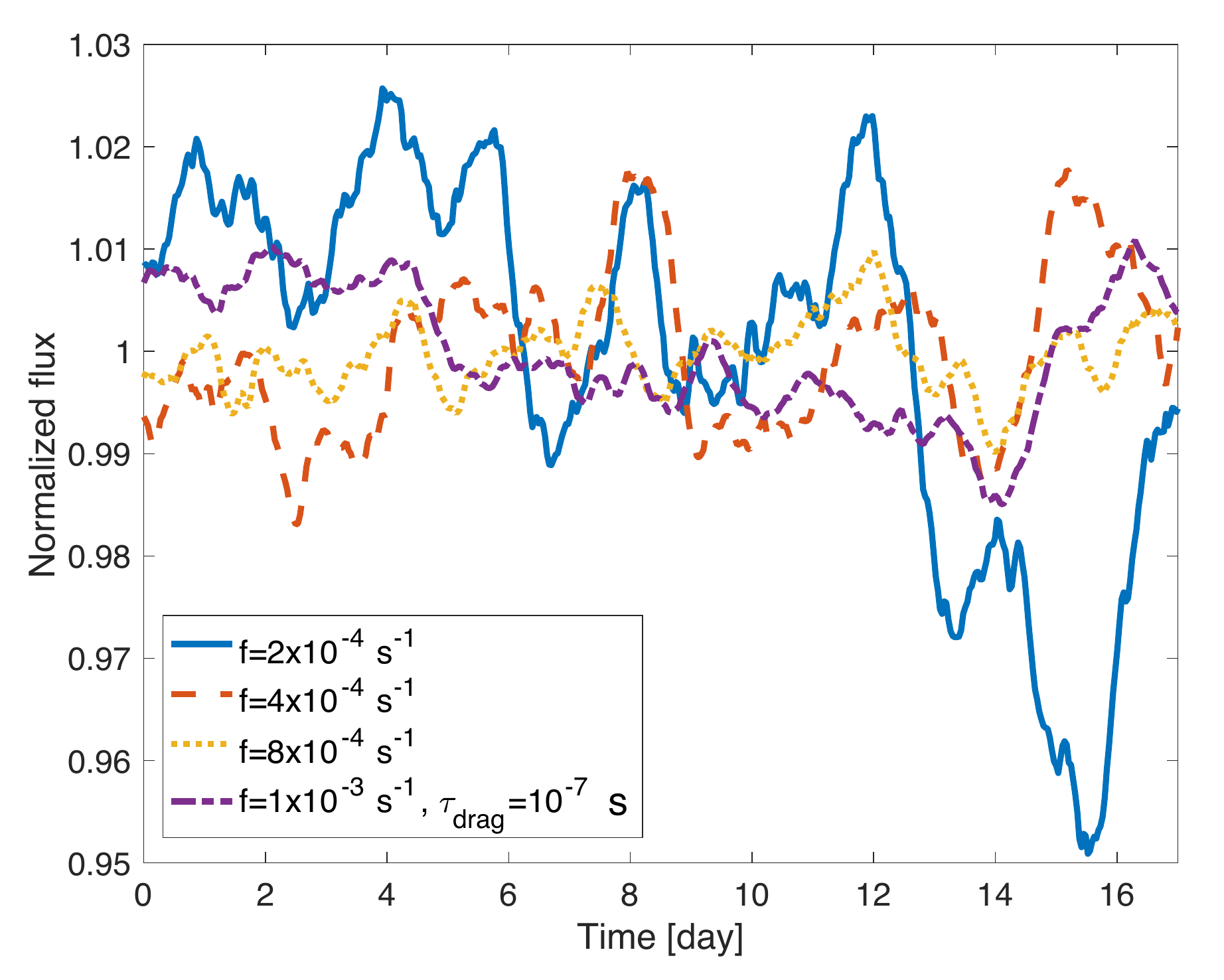}
    \caption{Normalized domain-averaged outgoing thermal flux as a function of time for {\btt three simulations with $\tdrag=10^5$ s and $f=2\times 10^{-4}, 4\times 10^{-4}$ and $8\times 10^{-4} \; \rm{s^{-1}}$ (corresponding to  $f$ at polar regions of objects with rotation periods of 17.5, 8.7 and 4.4 hours, respectively), and for a simulation with $\tdrag=10^7$ s and  $f=1\times 10^{-3}\;{\rm s^{-1}}$ (corresponding to $f$ at the polar region of an object with a rotation period of 3.5 hours).  All models are in a fixed domain size of $30000\;{\rm km}\times30000 \;\rm{km}$.} The lightcurves are normalized relative to their time-mean values.}
\label{fig.fluxvariability}
\end{figure}

Most observed lightcurve variability of BDs and free-floating EGPs  are thought to be caused by rotational modulation of surface inhomogeneity \citep{biller2017,artigau2018}. The variability often appears to be periodic, indicating the presence of quasi-stationary surface features. However, a fraction of  the variability  exhibits  changes over short timescales comparable to the rotation period, suggesting that the surface features evolve over short timescales \citep{artigau2009,wilson2014,metchev2015,apai2017}. A small fraction of them even show  long-period (over 20 hours), non-sinusoidal variability \citep{metchev2015}, and it is unclear whether rotational modulation is the typical cause.  The stochastic evolution of  storms driven by cloud radiative feedback {\btt may help} to explain these irregularities in the lightcurve variability. {\btt Figure \ref{fig.fluxvariability} shows the domain-averaged outgoing thermal flux as a function of time for three simulations with $\tdrag=10^5$ s and $f=2\times 10^{-4}, 4\times 10^{-4}$ and $8\times 10^{-4} \; \rm{s^{-1}}$, and for a simulation with $\tdrag=10^7$ s and $f=1\times 10^{-3} \; \rm{s^{-1}}$. } The  variability in these simulations are solely caused by the statistical fluctuation of the storm system.     The case with $f=2\times 10^{-4} \; \rm{s^{-1}}$ exhibits the largest  fractional flux variability up to about 7\% peak-to-peak variation, while that with $8\times 10^{-4} \; \rm{s^{-1}}$ is the smallest among these three cases (about 2\% peak-to-peak variation). {\btt Although the case with $\tdrag=10^7$ s and $f=1\times 10^{-3} \; \rm{s^{-1}}$ develops domain-size vortices (see Section \ref{ch.bottomdrag}), the amplitude of its resulting lightcurve variability is not obviously larger than that of other cases shown in Figure \ref{fig.fluxvariability}.  } The typical evolution timescale of the flux is tens of hours, longer than the overturning timescale of an individual storm (which can be roughly estimated as on the order of $\sim 10^4$ s from the dominant lengthscales and wind speed of the storms). Figure \ref{fig.fluxvariability} suggests that the more storms in the domain, the less effect that the fluctuation of a single storm would have in the  total flux variability.

In the planetary surface, the latitudinal-dependence  of the Coriolis parameter $f$ indicates latitudinal-dependent sizes of local storms. Near the equator where $f$ is small, storms might be able to grow to a lengthscale that is a nontrivial fraction of the planetary radius. For example, a BD with a 5-hour rotation  has a Coriolis parameter at $f=1.2\times 10^{-4} \; \rm{s^{-1}}$ at $10^{\circ}$ latitude and $f=7\times 10^{-4} \; \rm{s^{-1}}$ at the pole. Given our model conditions, this results in a typical storm diameter of about $1.8\times10^4$ km at $10^{\circ}$ latitude and $3\times10^3$ km on the poles. Assuming a Jupiter radius,  a single storm near the pole covers less than $\sim 0.1\%$ of the disk  whereas that at $10^{\circ}$ latitude covers $\sim 1.7\%$ of the disk.  If the BD or the directly imaged EGP is observed  pole on, statistical fluctuation of storms is expected to cause negligible flux variability. However, if the view  equator on, not only rotational modulation is maximized, but the evolution of the larger storms may induce additional variability over a timescale on the order of tens of hours. This {\btt may contribute to} the irregular variability of many observed lightcurves of BDs (e.g., \citealp{metchev2015}). 

When vertical transport of clouds is dominated by large-scale flows via cloud radiative feedback, one would expect that clouds are thicker at low latitudes and thinner at high latitudes for a BD with certain rotation rate. If the BD rotates sufficiently slow, the  equator-to-pole cloud thickness variation may be small.   Under similar atmospheric conditions --- including temperature, gravity and metallicity,  BDs that rotate faster are expected to have overall thinner clouds than those rotate slower. These consequences have  implications for observed near-IR colors for BDs and directly imaged EGPs as the thickness of clouds affects the spectral properties. For example, different overall cloud thickness caused by  different rotation rate may contribute to the near-IR color scattering of mid-to-late L dwarfs (e.g., \citealp{faherty2016}). Recently,  \cite{vos2017} and \cite{vos2020} suggested that BDs that are viewed from more equator-on tend to exhibit redder near-IR colors and larger flux variability than those viewed more pole-on, for which our numerical results naturally support.

\subsection{Unresolved issues  and outlook}

We have shown in Figure \ref{fig.patchiness} that cloud thickness decreases with increasing rotation rate. One would intuitively expect that  faster rotation can lead to a more stringent balanced state of the flow given  roughly the same amount of cloud radiative feedback, and therefore weaker vertical motions to transport tracers against particle settling.   However, this picture itself would ${\it not}$ automatically imply a mechanism. This is a central unresolved issue of this study. Revealing the detailed mechanism likely requires an understanding of the eddies and their effect on the mean flows (here, eddies refer to motions with scales smaller than the vortex scale, and the mean flows refer to the vortex motions). If the vortex is roughly circular, the flow  can be decomposed into axisymmetric and non-axisymmetric components. The axisymmetric flow will be largely in balance between the pressure gradient force and the Coriolis force (plus the centrifugal force if it is important), with a residual  component providing divergence and convergence of the vortex that are essential to drive cloud formation. The balance of the residual overturning circulation is primarily between angular momentum transport by eddies and the Coriolis force associated with  the  residual overturning flow if the vortex is close to quasi-geostrophic balance (e.g., \citealp{showman&kaspi2013}). Therefore, the cloud-forming overturning circulation is essentially both thermally driven and eddy regulated. 
Furthermore, near the cloud top, as we have shown in Figure \ref{fig.wqspectra_allpressure}, eddies contribute significantly to the net upward transport of clouds. Cloud vertical  transport near the cloud top is essential because the cloud-top pressure determines the outgoing thermal flux and therefore the strength of the  atmospheric heating. In summary, understanding eddies is essential to pin down the mechanism by which stronger rotation leads to vertically thinner  clouds.  Future models adopting various level of complexity are needed to investigate this mechanism.

{\btt Our models show sensitivity to the strength of prescribed bottom frictional drag. This is because kinetic energy generated in the cloud forming region (which lies well above the region where drag is applied) can be transferred to deeper layers, forming dynamical modes that tend to be pressure independent and with horizontal lengthscales larger than the deformation radius. The strength of this mode is sensitive to the rate of energy removed by the bottom drag  (see results and discussion in Section \ref{ch.bottomdrag}). This highlights the fact that conditions in  layers deeper than the observable layers may influence observable flow and cloud properties. The form and strength of the drag are highly uncertain due to the unknown nature of interactions between the interior and the simulated shallow outer layer.  To date, no systematic theoretical study provides a guidance as to how such interactions should be parameterized in shallow models.  The uncertainties in the bottom boundary conditions  and the impacts on the general circulation  have long been an  issue for shallow models of gaseous planets (e.g., \citealp{ showman2020}). Such issues have been shown for shallow models relevant to Jupiter (e.g., \citealp{schneider2009}) and to hot Jupiters (e.g., \citealp{mayne2017,sainsbury2019,carone2020}).  }

Our idealized models are dedicated to a clear understanding of dynamical mechanisms. An obvious next step within this framework is to extend the model domain to a global geometry which includes effects of the latitudinal variation of $f$---the so-called $\beta$ effect where $\beta=df/dy$. The $\beta$ effect plays a central role in driving zonal banding and jets in Jovian and Saturnian atmospheres (see reviews by, for example, \citealp{vasavada2005,showman2018review}).  This would help to clarify whether BDs and directly EGPs exhibit zonal jets like Jupiter and Saturn as indicated by long-term lightcurve monitoring \citep{apai2017} and simultaneous tracking of radio and near-IR flux variability \citep{allers2020}. {\btt Global models are also necessary to assess  whether circulation driven by cloud radiative feedback can quantitatively explain the observed lightcurve variability of BDs. } Future developments beyond  the current idealized framework include adopting a realistic radiative transfer scheme coupled to realistic chemistry and cloud formation---this allows direct comparisons between global models and the  rich observed spectrum, near-IR colors and spectral time variability of  BDs and directly EGPs. 

{\btt An alternative scenario associated with thermo-chemical effects of  ${\rm CO/CH_4}$ and ${\rm N_2/NH_3}$  conversion has been proposed to explain L dwarf spectrum and the L/T transition using static, 1D radiative-convective equilibrium framework \citep{tremblin2016,tremblin2019}. In our dynamical framework, cloud is chosen as a key dynamically active tracer for its excessive opacity in typical L and early T dwarfs \citep{marley2015}. When the chemical conversion timescales is close to or longer than dynamical timescales, the potential correlation between horizontal distribution of chemical species (therefore their radiative feedback) and large-scale dynamics may play a similar role as a dynamically active tracer, contributing to the heating/cooling rates that shape the circulation. Such radiative feedback by chemical quenching on the general circulation has been investigated for hot Jupiters (e.g., \citealp{steinrueck2019,drummond2020}), and will be an interesting topic in future studies for BDs and directly imaged EGPs. }


\section{Conclusions}
\label{ch.conclusion}

Cloud radiative feedback may play an essential role in driving  vigorous  circulation in atmospheres of self-luminous, substellar objects including  brown dwarfs (BDs) and directly imaged extrasolar giant planets (EGPs). In this study, we have numerically investigated the atmospheric circulation in conditions relevant to these objects using a general circulation model that is self-consistently coupled with idealized cloud formation and its radiative feedback. As a first step in this line of study and to better understand the effects of rotation on the turbulent cloud formation, our models adopt a constant Coriolis parameter $f$ across a Cartesian domain  and with a double-periodic horizontal boundary condition.
We have reached  the following key conclusions:

\begin{itemize}
    
    \item Vigorous atmospheric circulation  can be triggered and self-sustained by  cloud radiative feedback in conditions appropriate for BDs and directly imaged EGPs. In a constant-$f$ approximation, the circulation is dominated by turbulent vortices, with thick clouds forming in the anticyclones and thin clouds or clear sky in the cyclones. {\btt Horizontally averaged wind speed can be several hundred $\mps$ with local wind speeds exceeding $1000\mps$, and isobaric temperature differences can be over 100 K} for plausible physical parameters. Fractional outgoing thermal flux differences between cloudy and cloudless regions can be comparable to 1. This is a natural and robust mechanism to generate significant surface inhomogeneity that is responsible for the observed lightcurve variability of BDs and  EGPs.
    
    \item In the presence of strong rotation and strong deep frictional drag (which crudely represents interactions between the weather layer and the deep quiescent interior), the characteristic horizontal lengthscales of  dominant vortices are close to the deformation radius $L_d=c_g/f$ where $c_g$ is the phase speed of gravity waves. In the absence of rotation or when rotation is weak, the circulation is characterized by convergent and divergent flows with horizontal lengthscales regulated by the radiative timescale of the atmosphere.
    
    \item Stronger rotation leads to vertically thinner clouds, and the cloud thickness is greatest in the absence of rotation. Both the vortex-scale flow and smaller-scale inertia gravity waves contribute to the vertical transport of vapor and clouds. The mechanism by which rotation regulates the cloud thickness is likely related to interactions between vortices and smaller-scale flows.
    
    \item When the deep frictional drag is weak, there can be two distinctive modes of vortices. The first mode has a typical horizontal lengthscale close to the deformation radius and is driven directly by the geostrophic adjustment associated with cloud radiative feedback. The other mode has a strong pressure-independent component and can grow to a horizontal lengthscale  comparable to the simulated domain size. The large, domain-scale vortices could  affect  efficiency of  cloud formation and induce horizontal drifts of  cloud-forming vortices with a size close to the deformation radius.
    
    \item Storms driven by cloud radiative feedback evolve over timescales of several to tens of hours, and the statistical fluctuation of the ensemble of storms could induce  variability of the domain-mean flux. Such flux variability may help to explain the irregular lightcurve variability observed in a fraction of  BDs.  The change of cloud thickness with different Coriolis parameter $f$ indicates that the  cloud thickness may be different at different latitudes of the BDs or directly imaged EGPs. The global-mean cloud thickness might also be different for objects with different rotation rate. These may contribute to the observed scattered near-IR colors of dusty L dwarfs and the viewing angle dependent near-IR colors of BDs.
    
\end{itemize}

\section*{Acknowledgements}
We thank Ray Pierrehumbert and Yifan Zhou for helpful discussion and Tad Komacek for comments on the draft.  X.T. acknowledges support from the European community through the ERC advanced grant EXOCONDENSE (PI: R.T. Pierrehumbert). This work was completed with resources provided by the department of Physics at University of Oxford and the Lunar and Planetary Laboratory at University of Arizona. 

\section*{Data availability}
The data underlying this article will be shared on reasonable request to the corresponding author.




\bibliographystyle{mnras}
\bibliography{draft} 




\appendix

\section{Numerical tests}
\label{ch.resolution}

\begin{figure*}
	\includegraphics[width=2.\columnwidth]{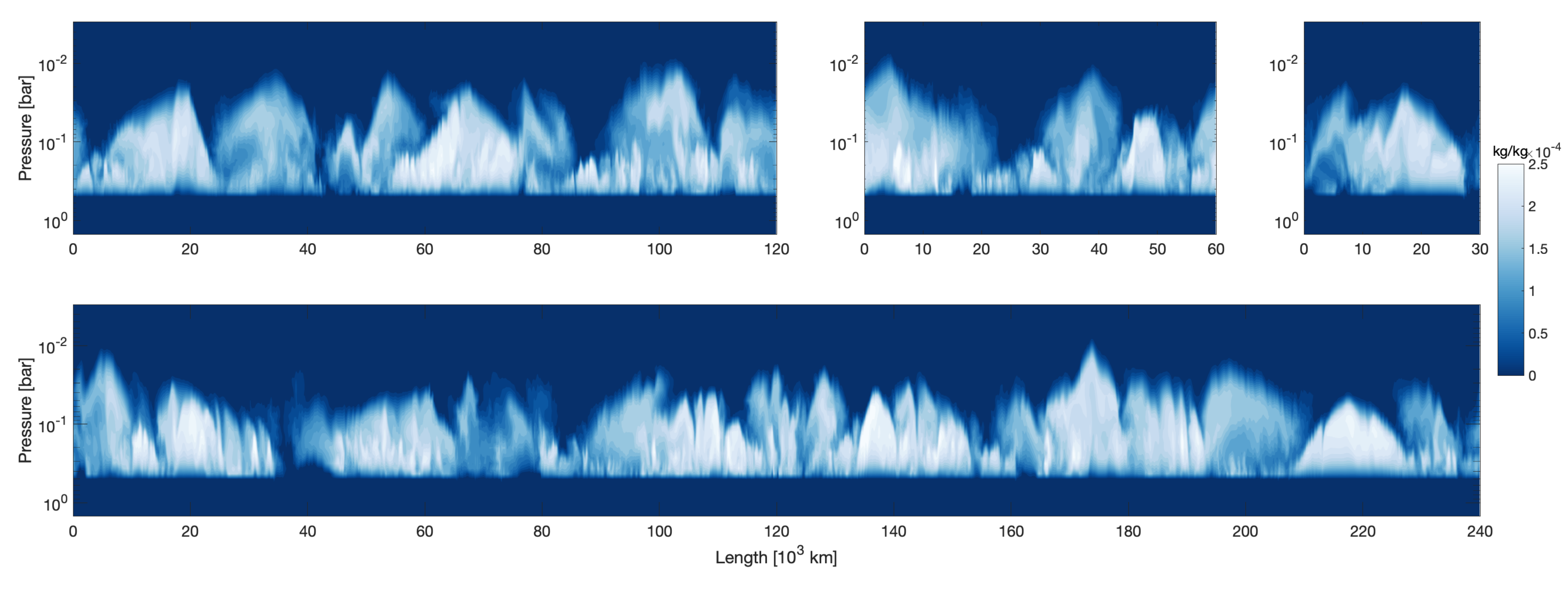}
    \caption{Instantaneous snapshots of cloud mixing ratio as a function of $x$ and pressure for 2D simulations with different domain size of $24\times 10^4$, $12\times 10^4$, $6\times 10^4$ and $3\times 10^4 \;\rm{km}$. Other model parameters, including the horizontal resolution of 150 km per grid cell, are the same among these simulations. }
\label{fig.domaintest.2d}
\end{figure*}

\begin{figure}
	\includegraphics[width=1.\columnwidth]{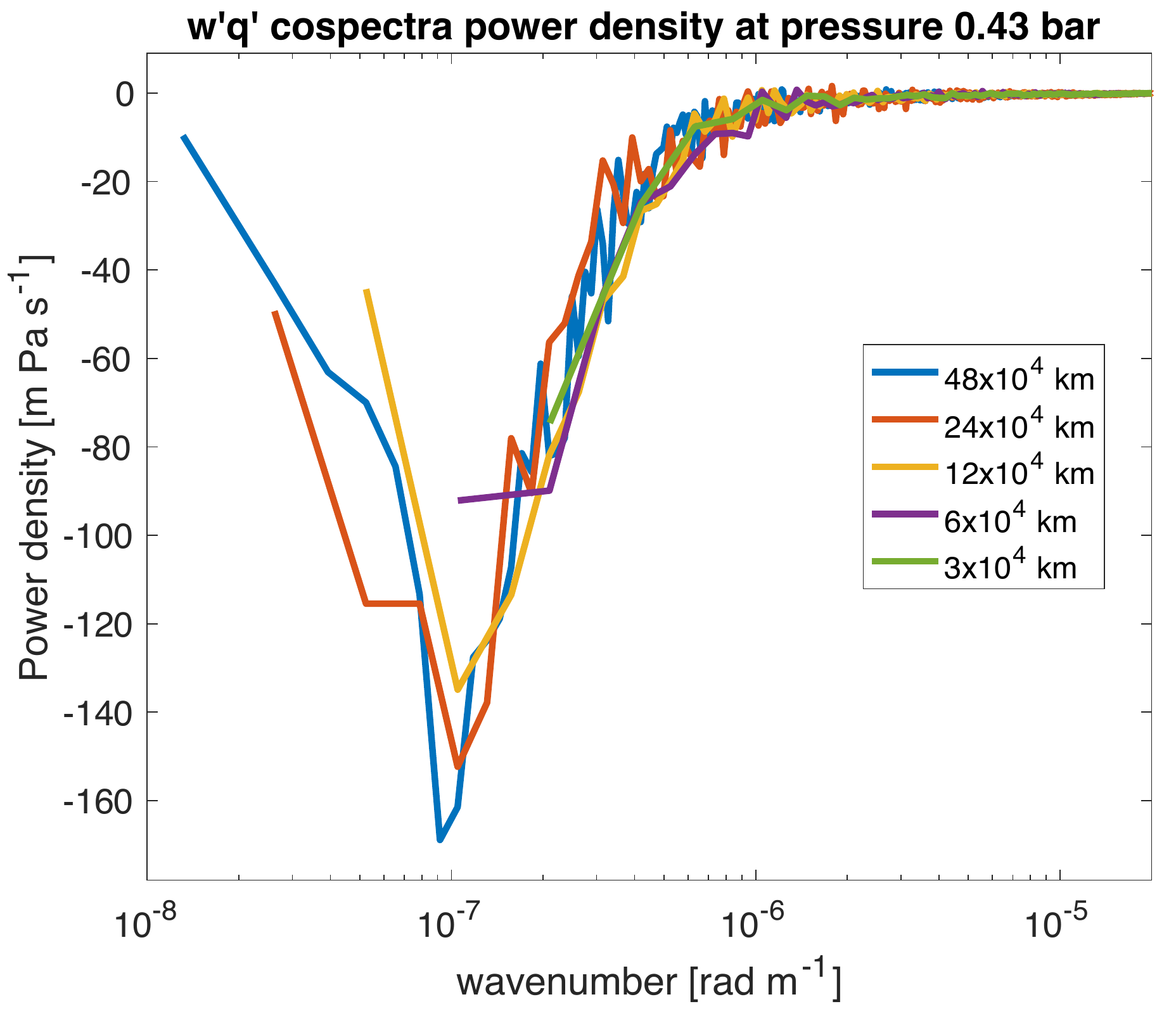}
    \caption{Cospectral power  density for $\omega' q'$ at 0.43 bar for 2D simulations shown in Figure \ref{fig.domaintest.2d} with an additional one with $48\times 10^4 \;\rm{km}$. }
\label{fig.wqspectrum_1d}
\end{figure}


 Figure \ref{fig.domaintest.2d} shows a test of the non-rotating 2D models with different domain length. Dominant aspects of the circulation, including the mean kinetic energy, mean cloud thickness and wind speeds are  not sensitive to the chosen domain size in Figure \ref{fig.domaintest.2d}. All simulations in Figure \ref{fig.domaintest.2d} exhibit  characteristic large-scale structures with a size of about $2\times 10^4 $ to $4\times 10^4$ km.
Figure \ref{fig.wqspectrum_1d} shows the cospectral power power density for $\omega' q'$ at 0.43 bar for the 2D simulations shown in Figure \ref{fig.domaintest.2d} with along an additional simulation with $48\times 10^4 \;\rm{km}$, demonstrating that flows with size $\sim 3\times 10^4$ km dominate the formation and vertical transport of clouds in the non-rotating 2D systems.


\section{Different forcing amplitudes}
\label{ch.forcing}

\begin{figure}
	\includegraphics[width=1.\columnwidth]{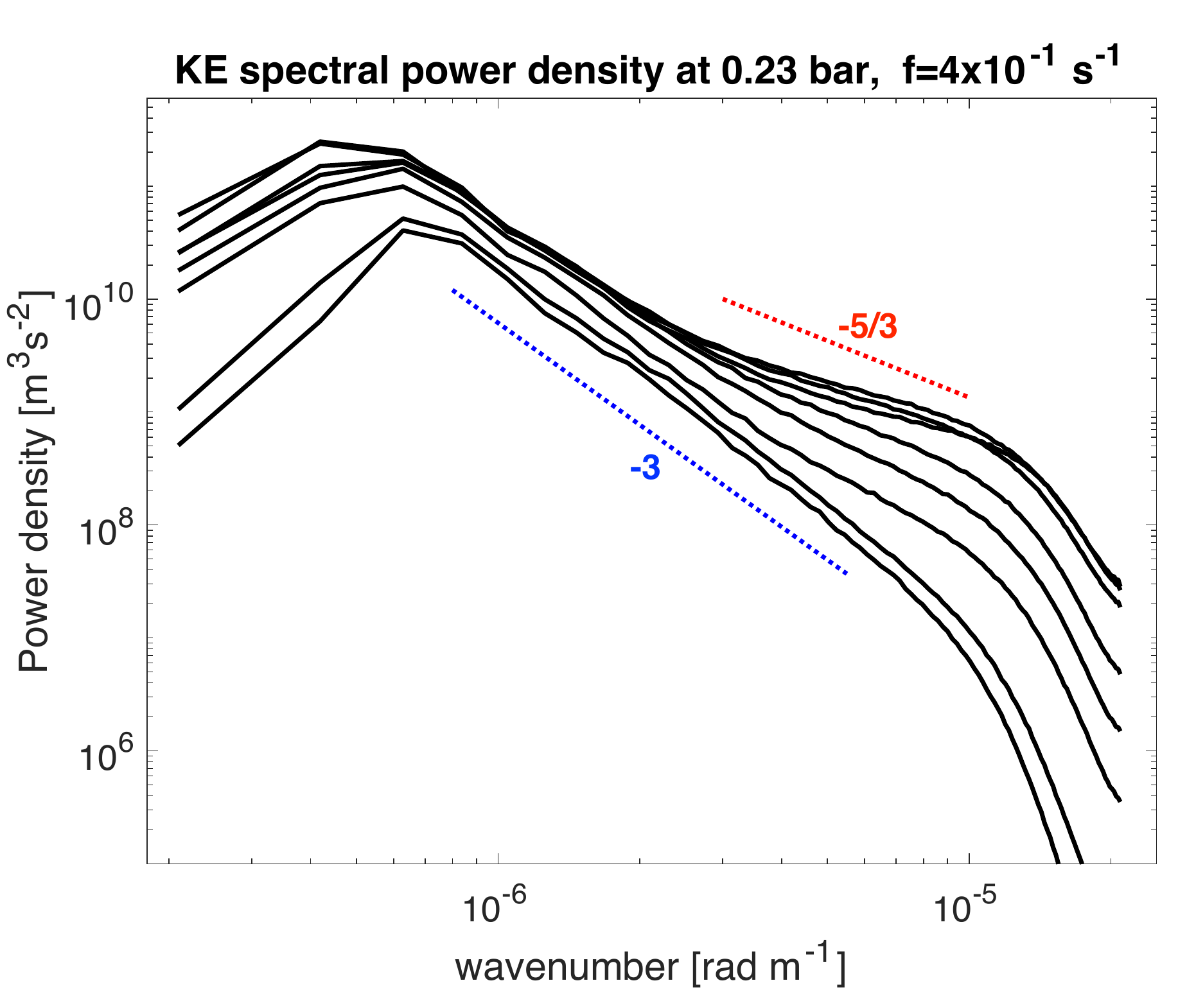}
    \caption{Kinetic energy power spectrum at a pressure of 0.23 bar from simulations  with different forcing amplitudes and a Coriolis parameter $f=4\times 10^{-4} \; \rm{s^{-1}}$. The overall KE spectral power increases as the forcing amplitude increases. }
\label{fig.spectrum_q}
\end{figure}

Simulations shown in Figure \ref{fig.spectrum_f} are somewhat strongly forced such that inertia gravity waves can be  energetically important in the power spectra at 0.23 bar when the wavenumber is sufficiently larger than that of the deformation radius.  We investigated effects of changing  the  forcing amplitude via adjusting the abundance of deep condensible vapor --- the smaller the condensible vapor, the less forcing. In practice the settling velocity of cloud particles is artificially adjusted smaller; this is because as the circulation weakens due to the smaller forcing, it may not be able to keep cloud particles aloft to sustain the circulation. Figure \ref{fig.spectrum_q} shows the KE power spectra for different forcing amplitudes. As the forcing becomes weaker, the transition from slope $-3$ to $-5/3$ gradually vanishes. In the two least forced cases,  their KE power spectra exhibit a slope of $-3$  from the deformation radius all the way down to the scale at which KE is affected by the numerical  dissipation, and no transition to $-5/3$ occurs. When the forcing is stronger, the KE power spectrum flattens out at relatively high wavenumber. Interestingly, the peak of the KE spectrum systematically moves to smaller wavenumber (larger size) as the forcing increases.


\bsp	
\label{lastpage}
\end{document}